\documentclass[aps,pra,reprint, amsmath, amssymb,floatfix,superscriptaddress]{revtex4-1}
 
\usepackage{bm}
\usepackage{bbm}
\usepackage[retainorgcmds]{IEEEtrantools}
\usepackage{graphicx}
\usepackage{mathrsfs}
\usepackage{amsmath}
\usepackage{amsfonts}
\usepackage{amssymb}
\usepackage{color,xcolor}
\usepackage{times}
\usepackage{nicefrac}
\usepackage{ragged2e}
\usepackage{float}
\usepackage{braket}
\usepackage{tikz}

\usepackage[colorlinks=true,linkcolor=blue,urlcolor=blue,citecolor=blue,pdfusetitle]{hyperref}

\usepackage{blindtext}
\usepackage{stmaryrd} 
\usepackage{dsfont}
\usepackage{amsthm}

\newcommand{\tr}{\mathrm{Tr}}

\newtheorem{result}{Result}

\begin{document}

\title{Deterministic Equations for Feedback Control of Open Quantum Systems III:\\ Full counting statistics for jump-based feedback}
\date{\today}
\author{Alberto J. B. Rosal}
\email{abezerra@ur.rochester.edu}
\affiliation{Department of Physics and Astronomy, University of Rochester, Rochester, New York 14627, USA}
\affiliation{University of Rochester Center for Coherence and Quantum Science, Rochester, New York 14627, USA}
\author{Guilherme Fiusa}
\affiliation{Department of Physics and Astronomy, University of Rochester, Rochester, New York 14627, USA}
\affiliation{University of Rochester Center for Coherence and Quantum Science, Rochester, New York 14627, USA}
\author{Patrick P. Potts}
\affiliation{Department of Physics and Swiss Nanoscience Institute,
University of Basel, Klingelbergstrasse 82 CH-4056, Switzerland}
\author{Gabriel T. Landi}
\affiliation{Department of Physics and Astronomy, University of Rochester, Rochester, New York 14627, USA}
\affiliation{University of Rochester Center for Coherence and Quantum Science, Rochester, New York 14627, USA}

\begin{abstract}

In this work, we consider a general feedback protocol based on quantum-jump detections, where the last detected jump channel is stored in a memory and subsequently used to implement a feedback action, such as modifying the system Hamiltonian conditioned on the last jump.
We show that the time evolution of this general protocol can be described by a Lindblad master equation defined in a hybrid classical–quantum space, where the classical part encodes the stored measurement record (memory) and the quantum part represents the monitored system.
Moreover, we show that this new representation can be used to fully characterize the counting statistics of a system subject to a general jump-based feedback protocol. 
We apply the formalism to a three-level system coupled to two thermal baths operating as a thermal machine, and we show that jump-based feedback can be used to convert the information obtained from the jump detections into work.
Our framework provides analytical tools that enable the characterization of key statistical properties of any counting observable under jump-based feedback, such as the average current, noise, correlation functions, and power spectrum.
\end{abstract}

\maketitle{}

\section{Introduction}
\label{sec:int}

A quantum system governed by a Markovian Master Equation can be dynamically described by a \textit{quantum trajectory} \cite{Tutorial,PhysRevB.63.125326,PhysRevLett.112.170401}. The system evolves through continuous no-jump dynamics interrupted by discrete transitions, or \textit{jumps}, where both the jump channels and their occurrence times are random variables \cite{PhysRevA.111.042415,PhysRevB.104.195408,PhysRevLett.127.198101,PhysRevB.107.245406}. 
These jumps may correspond, for example, to transitions between energy levels, as observed in photon-counting experiments~\cite{PhysRevLett.57.1699,PhysRevLett.57.1696,PhysRevLett.56.2797,PhysRevLett.54.1023}.
A natural strategy is to detect quantum jumps in order to acquire information about the system, and then use this information to implement feedback protocols \cite{Rosal2025DeterministicFeedback,Wiseman1994_Feedback}. 
Feedback enables key quantum tasks such as cooling~\cite{Rosal2025MemoryStatFB,PhysRevLett.90.043001,PhysRevA.74.012322,PhysRevLett.96.043003,PhysRevA.91.043812,PhysRevLett.117.163601,PhysRevLett.123.223602,PhysRevLett.122.070603,PhysRevA.107.023516,CoolingGui,PhysRevB.68.235328,PhysRevLett.119.123603,PhysRevA.98.023828}, error correction~\cite{QECwithFeedback,QuantumThermodynamicsForQuantumComputing}, coherent control~\cite{Vijay2012,DiscreteFeedback2,PhysRevA.67.052101,PhysRevLett.120.073601}, and work extraction~\cite{QuantumDemon1,QuantumDemon2,QuantumDemon3,SecondLawFeedback1,SecondLawFeedback2,SecondLawFeedback3}, making it central to both theoretical and experimental advances in modern quantum platforms~\cite{Minev2019CatchingReverseQuantumJump,PhysRevLett.117.206803,PhysRevApplied.23.044063}.

As shown by Wiseman~\cite{Wiseman1994_Feedback}, one can consider the following feedback protocol: whenever a jump is detected, an instantaneous quantum channel is applied to the system.
However, most of the time the system undergoes no-jump evolution~\cite{Tutorial}, during which no feedback is applied. Moreover, this strategy does not retain information about previously detected jumps.
In contrast, Ref.~\cite{Rosal2025DeterministicFeedback} introduced a general feedback protocol that stores the last jump in a memory, which is then used to determine the feedback action.
The key distinction is that the feedback does not need to be applied immediately after the jump; rather, it may be triggered during the no-jump evolution.
For example, conditioned on the last jump channel, one may adjust the system Hamiltonian, such as by tuning energy gaps or applying an external drive, thus using the jump information to modify the no-jump evolution.
Such strategies have direct experimental realizations in quantum dot setups~\cite{PhysRevLett.117.206803,PhysRevApplied.23.044063}, and are not encompassed by Ref.~\cite{Wiseman1994_Feedback}, though fully described by Ref.~\cite{Rosal2025DeterministicFeedback}.

Full counting statistics (FCS) quantifies the probability distribution of the number of quantum jumps over time~\cite{Tutorial}.
We introduce the \emph{stochastic charge}~\cite{Tutorial,FPTLandi} (also referred to as the \emph{counting observable}~\cite{Fiusa2025CountingObservables,Fiusa2025FrameworkFluctuatingTimes}) as
\begin{equation}
\label{eq: stoc charge int}
    N(t) = \sum_{k\in\Sigma} \nu_k N_k(t)~,
\end{equation}
where $\Sigma$ denotes the set of experimentally monitored jump channels, and $N_k(t)$ is the (random) number of jumps in channel $k$ over the interval $[0,t]$.
By appropriately defining the weights $\nu_k$, different observables can be extracted, such as entropy production~\cite{Fiusa2025CountingObservables,Fiusa2025FrameworkFluctuatingTimes} and dynamical activity~\cite{PhysRevResearch.5.023155,PhysRevE.109.044114}.

In this work, we present a general framework to describe the statistical properties of a counting observable $N(t)$ under feedback conditioned on the last jump channel, thereby linking jump-based feedback dynamics with full counting statistics.
It has important implications for quantum thermodynamics, particularly for the implementation of Maxwell demons via quantum feedback~\cite{QuantumDemon1,QuantumDemon2,QuantumDemon3}, and for feedback-controlled systems based on jump monitoring~\cite{Sayrin2011RealTimeFeedback,Minev2019CatchingReverseQuantumJump,PhysRevLett.117.206803,PhysRevApplied.23.044063}.

This work is organized as follows.
In Sec.~\ref{sec:form_FCS}, we introduce the formalism of quantum jump detections; Sec.~\ref{subsec: Jump memory and memory-resolved state} presents the framework for jump-based feedback dynamics developed in Refs.~\cite{Rosal2025MemoryStatFB,Rosal2025DeterministicFeedback}; and in Sec.~\ref{subsec:form_simple_example}, we illustrate two simple examples of jump-based feedback.
Further discussion of the properties and interpretation of the resulting jump-based master equation is provided in Sec.~\ref{subsec: interpretation of the jump-based ME}, and a comparison with earlier jump-based feedback protocols is given in Sec.~\ref{subsec: previous jump fb}.
In Sec.~\ref{sec:results}, we present our first main result, showing that the time evolution of a general jump-based feedback protocol can be equivalently described by a Markovian Master Equation in an extended hybrid space, consisting of a quantum subsystem (the system under jump monitoring) and a classical subsystem (the memory storing the last detected jump).
The second main result demonstrates how this new feedback representation can be used to describe the statistical properties of a stochastic charge $N(t)$ [Eq.~\eqref{eq: stoc charge int}] for general jump-based feedback protocols.
In Sec.~\ref{sec:applicatio_maser}, we apply this framework to a three-level system coupled to two thermal baths. We show that a jump-based feedback protocol can be used to enforce the system to perform work on the environment regardless of the bath temperatures, effectively converting information from the jumps into work.
For this example, we also illustrate how to compute both the power and the fluctuations of the extracted work.
Finally, in Sec.~\ref{sec:conclusion}, we conclude with a discussion of the broader implications of our results.

\section{Jump-based feedback master equation}
\label{sec:formalism}

\subsection{Quantum jumps and full counting statistics}
\label{sec:form_FCS}
Let us consider a system that evolves according to a quantum master equation 
\begin{equation}
\label{QME}
 \partial_t \rho_t =\mathcal{L}\rho= -i \big[H,\rho_t\big] +  \sum_{k\in\Sigma} \mathcal{D}[L_k]\rho_t\end{equation}
with Hamiltonian $H$ and jump operators $L_k$, where $\Sigma$ is the set of possible jumps, and $\mathcal{D}[L]\rho\equiv L\rho L^\dagger - 1/2 \{L^\dagger L,\rho\}$ denotes the dissipator. 
Equation~\eqref{QME} describes the continuous-time evolution of the state $\rho_t$ in the system’s state space.
Alternatively, it can be expressed as a discrete evolution from $\rho_t$ to $\rho_{t+\delta t}$ over an infinitesimal time step $\delta t > 0$, as given by~\cite{Tutorial}
\begin{equation}
\label{eq: unraveling}
    \rho_{t+\delta t} = V_0 \rho_t V_0^\dagger + \sum_{k\in\Sigma} V_k \rho_t V_k^\dagger + \mathcal{O}(\delta t^2)~,
\end{equation}
where we retain terms only up to first order in $\delta t$.
Here, $V_0 \equiv \mathrm{1} - i\delta t\, H_{\text{eff}}$, with the effective non-Hermitian Hamiltonian $ H_{\text{eff}} \equiv H - \frac{i}{2}\sum_{k\in\Sigma} L_k^\dagger L_k$, and $V_k \equiv \sqrt{\delta t}\, L_k$ for $k\in\Sigma$. One can verify that
\begin{equation}
    V_0^\dagger V_0 + \sum_{k\in\Sigma} V_k^\dagger V_k = \mathrm{1} + \mathcal{O}(\delta t^2)~,
\end{equation}
where $\mathrm{1}$ is the identity operator.
Thus, for infinitesimal $\delta t$, the set $\{V_k\}$ constitutes a valid set of Kraus operators. 
The decomposition of the master equation~\eqref{QME} into this infinitesimal form [Eq.~\eqref{eq: unraveling}] is known as the \emph{unraveling of the master equation}, and defines the \emph{quantum-jump Kraus operators}.
The operator $V_0$ describes the no-jump evolution, whereas $V_k$ ($k \in \Sigma$) represents a jump through channel $k$.

For some systems, the set of Kraus operators $\{V_k\}_{k=0, k\in\Sigma}$ defines an accessible measurement scheme, often realized through photon-counting experiments~\cite{PhysRevLett.57.1699,PhysRevLett.57.1696,PhysRevLett.56.2797,PhysRevLett.54.1023}.
Continuous monitoring of these jumps yields an outcome $x_t$ at time $t$, where $x_t = 0$ denotes a no-jump detection associated with $V_0$, and $x_t = k $ corresponds to a jump in channel $k$ described by $V_k$.
The (quantum) stochastic trajectory up to time $t$ is then represented by the dataset $\Gamma_t = \{x_{t'}\}_{t' \in [0,t] }$.
When we detect $x_t = 0$, the system undergoes a no-jump evolution generated by the no-jump Liouvillian $\mathcal{L}_0$. 
In contrast, when a jump is detected, $x_t = k$, the system evolves according to the jump channel $\mathcal{J}_k$. 
These dynamical maps are defined as
\begin{eqnarray}
\label{eq: jump channel superoperator}
    \mathcal{J}_k\rho &\equiv& V_k \rho V_k^\dagger~,\\
\label{eq: no jump liovillian}
    \mathcal{L}_0 &\equiv& \mathcal{L} - \sum_{k\in\Sigma} \mathcal{J}_k~,
\end{eqnarray}
where we can show that $(\mathrm{1} + \delta t~ \mathcal{L}_0) \rho = V_0 \rho V_0^\dagger$.

By assigning weights $\nu_k$ to each channel $k$ in the counting observable $N(t)$ [Eq.~\eqref{eq: stoc charge int}], one can capture all the key observables of the system. 
For instance, if jump $k$ corresponds to an energy-level transition induced by a thermal bath at inverse temperature $\beta_k$, and $\Delta E_k$ denotes the associated energy change, one can set $\nu_k = - \beta_k \Delta E_k$, so that the $N(t)$ represents the entropy production of the system over the interval $[0,t]$.
Furthermore, by choosing $\nu_k = 1$ for all $k$, one obtains the \emph{dynamical activity}, which counts the total number of jumps irrespective of their channels ~\cite{PhysRevResearch.5.023155,PhysRevE.109.044114}.
In Sec.~\ref{sec:applicatio_maser}, we present an explicit example in which $N(t)$ corresponds to the work performed by a quantum thermal machine.
Consequently, the statistics of $N(t)$ play a fundamental role in describing physical systems~\cite{Fiusa2025CountingObservables,Fiusa2025FrameworkFluctuatingTimes}.

\subsection{ Jump memory and memory-resolved state}
\label{subsec: Jump memory and memory-resolved state}

The \textit{jump memory} $k_t$~\cite{Rosal2025DeterministicFeedback} is defined as the stochastic process that records the last detected jump of the quantum trajectory $\Gamma_t$. For instance, if one has $\Gamma_t = \{\cdots0,0,q,0,0,0,0,\bar{k},0,0\}$, then $k_t = \bar{k}$. 
A general jump-based feedback is then implemented if we use the information of the last jump encoded in $k_t$ to modify the system's Hamiltonian \emph{and/or} jump operators. 
In other words, the system evolves under a stochastic Hamiltonian $H(k_t)$ and a set of jump operators $\{L_q(k_t)\}_{q\in\Sigma}$, both of which depend on the last jump channel recorded by $k_t$.

Ref.~\cite{Rosal2025DeterministicFeedback} developed a framework that describes a general feedback protocol using deterministic equations.  
To apply this formalism, we first introduce the \textit{memory-resolved state},
\begin{equation}
    \varrho_t(k) \equiv E[\rho^c_t ~ \delta_{k,k_t}]~,
\end{equation}
where $\rho^c_t$ is the conditional state associated with a stochastic trajectory $\Gamma_t$, $E[\cdot]$ denotes the average over all possible trajectories, $\delta_{a,b}$ is the Kronecker delta, and $k$ represents a possible realization of the jump memory $k_t$.
The probability $P_t(k)$ that $k$ is the last detected jump at time $t$ (i.e., the event $k_t = k$), as well as the unconditional quantum state $\bar{\rho}_t$, are obtained from $\varrho_t(k)$ as
\begin{eqnarray}
    \label{eq: memory distribution from SRR state}
     P_t(k) =    \mathrm{Tr}[\varrho_t(k)]~,~\bar{\rho}_t =\sum_{k}\varrho_t(k)~.
\end{eqnarray}
It was shown in Ref.~\cite{Rosal2025DeterministicFeedback} (see Appendix~\ref{ap: jump fb dyn} for an alternative proof) that, for a general feedback dynamics with a stochastic Hamiltonian $H(k_t)$ and jump operators $\{L_q(k_t)\}_{q\in\Sigma}$, the state $\varrho_t(k)$ evolves in time according to
\begin{eqnarray}
\label{eq: jump FB dyn}
    \partial_t \varrho_t(k) &=& -i \big[H(k),\varrho_t(k)\big] - \frac{1}{2} \sum_{q \in \Sigma} \big\{ L_{q}^\dagger(k)L_{q}(k),\varrho_t(k)\big\} \nonumber \\
                             & &  +\sum_{q \in \Sigma}\ L_k^{}(q) \varrho_t(q) L_k^\dagger(q).
\end{eqnarray}

Equation~\eqref{eq: jump FB dyn} is referred to as the \emph{jump-based feedback master equation}.
In general, once the dependence of the Hamiltonian $H(k)$ and the jump operators $L_{q}(k)$ on each jump channel $k$ is specified, Eq.~\eqref{eq: jump FB dyn} can be solved to obtain the full feedback dynamics, yielding both the jump-memory probability distribution $P_t(k)$ and the unconditional system state $\bar{\rho}_t$ [Eq.~\eqref{eq: memory distribution from SRR state}].
The state $\varrho_t(k)$ describes two degrees of freedom: one is classical (corresponding to the probability distribution of the memory $k_t$), and the other is quantum (representing the quantum state $\bar{\rho}_t$ of the system). 
In what follows, we present simple examples of jump-based feedback protocols that illustrate the type of feedback dynamics described by Eq.~\eqref{eq: jump FB dyn}.

\subsection{Simple examples}
\label{subsec:form_simple_example}

A feedback strategy based on the last detected jump may consist of turning on an external drive in the Hamiltonian conditioned on the last jump channel $k_t$~\cite{Rosal2025DeterministicFeedback}. 
Consider a two-level system (qubit) with computational basis states $\ket{g}$ (ground) and $\ket{e}$ (excited), separated by an energy gap $\omega$, and coupled to a thermal bath. 
In the absence of feedback, the system dynamics is governed by the master equation~\eqref{QME} with jump operators $ L_- = \sqrt{\gamma_-}\,\ket{g}\!\bra{e}$ and $L_+ = \sqrt{\gamma_+}\,\ket{e}\!\bra{g}$, which describe, respectively, the emission and absorption of a thermal photon by the qubit. 
Thus, the system admits two possible transitions, where we denote $k=+1$ for absorption and $k=-1$ for emission.

The incoherent jump $\ket{e}\to\ket{g}$ can be monitored by detecting the photon emitted into the thermal environment, as experimentally demonstrated through the observation of intermittent fluorescence in trapped ions~\cite{PhysRevLett.57.1699,PhysRevLett.57.1696,PhysRevLett.56.2797,PhysRevLett.54.1023}. In addition, the jump $\ket{g}\to\ket{e}$ can be detected by continuously monitoring the excited state, as demonstrated in superconducting artificial three-level atoms~\cite{Minev2019CatchingReverseQuantumJump}.
Assuming continuous monitoring of such transitions, the quantum jump detection outcome $x_t$ can take the values $x_t = 0$ for no-jump detections or $x_t = \pm 1$ for emissions or absorptions at time $t$. 
The memory $k_t$ records the last transition experienced by the qubit and therefore takes values in $\{\pm 1\}$.

One can consider the following protocol: if the last transition was an absorption ($k_t = +1$), the external drive is turned on; if it was an emission ($k_t = -1$), the drive is removed. 
This results in the stochastic Hamiltonian
\begin{equation}
    H(k_t) = -\frac{\Delta}{2}\sigma_z + \lambda\,\delta_{k_t, 1}\,\sigma_x~,
\end{equation}
where $\lambda$ is the drive strength, $\Delta$ is the detuning, and $\sigma_{x,y,z}$ are the Pauli matrices.
In our convention, $\sigma_z = \ket{g}\bra{g} - \ket{e}\bra{e}$, so that the ground state $\ket{g}$ is the eigenstate of $\sigma_z$ with eigenvalue $1$.
At time $t$ one has either $H(-1) = -\frac{\Delta}{2}\sigma_z$ or $H(1) = -\frac{\Delta}{2}\sigma_z + \lambda \sigma_x$, corresponding to the two possible realizations of $k_t$. 
Note that this protocol does not modify the jump operators, as it only adds a term to the qubit Hamiltonian.

For this example, Eq.~\eqref{eq: jump FB dyn} becomes
\begin{eqnarray}
\label{eq: simple example 1}
    \partial_t \varrho_t(1)&= -i \big[H_0+\lambda \sigma_x,\varrho_t(1)\big]- \frac{1}{2}   \big\{ L,\varrho_t(1)\big\}\nonumber\\
                              & +\ L_+^{}\left( \varrho_t(1)+\varrho_t(-1)\right) L_+^\dagger~,
\end{eqnarray}

\begin{eqnarray}
\label{eq: simple example -1}
\partial_t \varrho_t(-1)&= -i \big[H_0,\varrho_t(-1)\big]- \frac{1}{2}   \big\{ L,\varrho_t(-1)\big\}\nonumber\\
                              & +\ L_-^{}\left( \varrho_t(1)+\varrho_t(-1)\right) L_-^\dagger~,
\end{eqnarray}
where $L \equiv L_+^\dagger L_++  L_-^\dagger L_- = \gamma_+ \ket{g}\bra{g} + \gamma_- \ket{e}\bra{e}$.
For the thermal environment coupling, we have $\gamma_+ = \gamma \bar{n}$ and $\gamma_- = \gamma (\bar{n}+1)$, where $\gamma$ is the coupling strength and $\bar{n} = (e^{\omega/T}-1)^{-1}$ is the Bose-Einstein occupation number of the bath at temperature $T$, with $\hbar = k_B = 1$.
These two coupled equations can be solved to obtain the memory-resolved states $\varrho_t(k)$ for $k = \pm 1$, yielding both the probabilities $P_t(k) = \tr[\varrho_t(k)]$ and the unconditional system state $\bar{\rho}_t = \varrho_t(1) + \varrho_t(-1)$.
In particular, the feedback steady-state can be obtained by setting $\partial_t \varrho_t(k) = 0$.

\begin{figure}
    \centering
    \includegraphics[width=\linewidth]{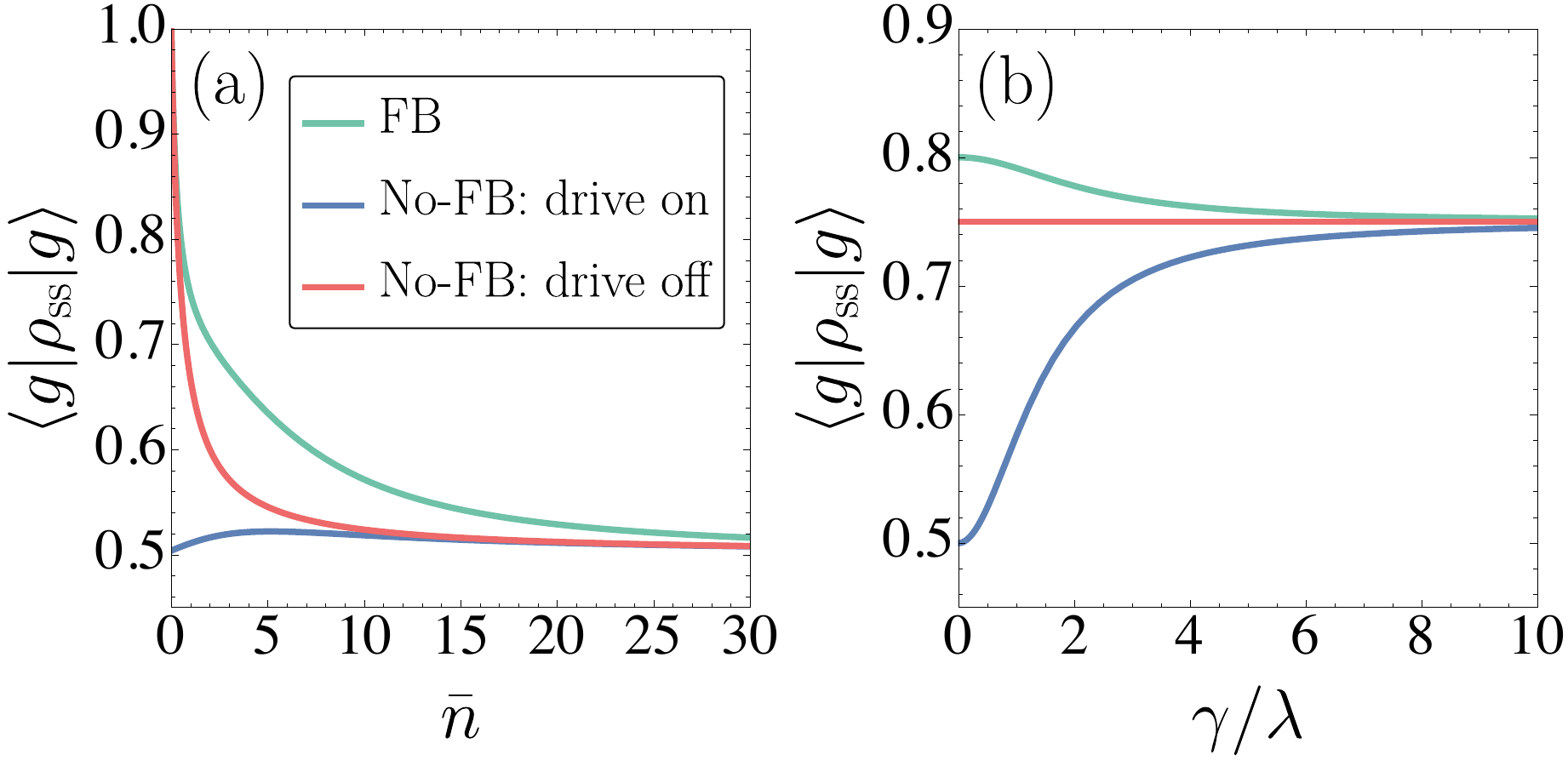}
    \caption{
    {} Population of the qubit's ground state in the stationary regime considering a resonant external drive ($\Delta = 0$). 
    (a) Dependence on the bath temperature. The feedback protocol increases the ground-state population, effectively implementing a cooling mechanism. Here, $\gamma/\lambda = 0.25$. 
    (b) Competition between thermal coupling $\gamma$ and drive strength $\lambda$. For strong drives ($\lambda \gg \gamma$), the cooling protocol becomes more efficient, further increasing the ground-state population. In this case, one has $\bar{n} = 0.5$.
  }
    \label{fig: example plots}
\end{figure}

In the stationary regime, the population of the ground state, $P_g \equiv \bra{g}\bar{\rho}_{\text{ss}}\ket{g}$, and the coherence, $C \equiv \bra{g}\bar{\rho}_{\text{ss}}\ket{e}$, are given by

\begin{eqnarray}
    P_g &=& \frac{(1+2\bar{n})(4+\bar{n}(1+\bar{n})(\gamma/\lambda)^2)}{4+\bar{n}(12+(1+2\bar{n})^2(\gamma/\lambda)^2)}~,\\
    C &=& -\frac{2i\bar{n}^2 (\gamma/\lambda)}{4+\bar{n}(12+(1+2\bar{n})^2(\gamma/\lambda)^2)}~,
\end{eqnarray}
where the population of the excited state in the stationary regime is $P_e = 1-P_g$. 
Furthermore, the distribution of the jump memory $k_t$ in the stationary regime ($t\to\infty$) is given by 
\begin{equation}
    P_{\text{ss}}(-1) = \frac{(1+\bar{n}) \bigl[4+\bar{n}(1+2\bar{n})(\gamma/\lambda)^2\bigr]}{4+\bar{n}\bigl[12+(1+2\bar{n})^2(\gamma/\lambda)^2\bigr]}~,
\end{equation}
where $P_{\text{ss}}(-1)$ is the probability that the last transition corresponds to an emission ($\ket{e}\to\ket{g}$), i.e., $\lim_{t\to\infty} k_t = -1$, whereas $P_{\text{ss}}(+1) = 1- P_{\text{ss}}(-1)$ is the probability that the last transition was an absorption ($\ket{g}\to\ket{e}$).

Figure~\ref{fig: example plots} shows the ground-state population under the feedback protocol in the stationary regime, compared with no-feedback cases where the external drive is always on or always off. 
In Fig.~\ref{fig: example plots}(a), $P_g = \bra{g}\bar{\rho}_{\text{ss}}\ket{g}$ is shown as a function of the bath's Bose--Einstein distribution $\bar{n}$, illustrating the protocol's performance across different temperatures.
The feedback effectively cools the qubit: upon detecting an excitation, the external drive returns the qubit to the ground state, thereby increasing its population relative to the no-feedback scenarios.
Figure~\ref{fig: example plots}(b) illustrates the competition between the drive strength $\lambda$ and the thermal coupling $\gamma$ under the feedback protocol.
In the weak-drive regime ($\gamma \gg \lambda$), all cases behave similarly, as the drive is too weak to overcome thermal dissipation and the feedback has little effect. 
In contrast, in the strong-drive regime ($\lambda \gg \gamma$), the feedback efficiently drives the system from $\ket{e}$ to $\ket{g}$ after an excitation, enhancing the ground-state population.

In this example, the feedback modifies only the qubit's Hamiltonian. However, one can also design feedback strategies that adjust the jump operators depending on the last detected transition. 
For instance, the energy gap of the system could be modified~\cite{PhysRevLett.117.206803}. For a qubit coupled to a thermal bath, the jump operators are 
$L_- = \sqrt{\gamma (\bar{n}+1)}\,\ket{g}\bra{e}$ and 
$L_+ = \sqrt{\gamma \bar{n}}\,\ket{e}\bra{g}$, where $\bar{n} = [\exp(\omega/T)-1]^{-1}$. 
Thus, by changing the energy gap $\omega$ based on the last detected jump $k_t$, both the system Hamiltonian and the corresponding jump operators are modified.
In this case, the system Hamiltonian can be written as $H(k_t) = -\frac{\omega(k_t)}{2}\sigma_z$, where the energy gap $\omega(k_t)$ depends on the last detected transition. 
Since the jump operators depend on $\bar{n}$ (and thus on $\omega$), they also acquire a dependence on $k_t$, giving $L_-(k_t)$ and $L_+(k_t)$. 
This type of feedback can be implemented experimentally by applying voltage gates to the qubit conditioned on the last detected jump, thereby modifying the energy gap between the ground and excited states, or equivalently via a quantum point contact~\cite{PhysRevLett.117.206803,PhysRevApplied.23.044063}.

\subsection{Interpretation of the jump-based feedback master equation}
\label{subsec: interpretation of the jump-based ME}
In general, it is not possible to write a closed master equation for the system state $\bar{\rho}_t$ under jump-based feedback dynamics.
For instance, summing over $k$ on both sides of Eq.~\eqref{eq: jump FB dyn} yields $\partial_t \bar{\rho}_t$ on the left-hand side. However, when the Hamiltonian $H(k)$ and/or the jump operators $L_q(k)$ depend on $k$, one can no longer obtain a closed equation for $\bar{\rho}_t$ on the right-hand side, since terms such as $\sum_k [H(k), \varrho_t(k)]$ couple the evolution of $\bar{\rho}_t$ to the memory-resolved states $\varrho_t(k)$.

On the other hand, Eq.~\eqref{eq: jump FB dyn} can be interpreted as a rate equation. Using the definition of $V_0$ introduced in Sec.~\ref{sec:form_FCS}, together with $(\mathrm{1}+\delta t\,\mathcal{L}_0)\rho = V_0 \rho V_0^\dagger$, one finds
\begin{equation}
\label{eq: no-jump liovilian and hamiltonian}
    \mathcal{L}_0(k) \varrho_t(k) = -i \big[H(k),\varrho_t(k)\big] - \frac{1}{2} \sum_{q \in \Sigma} \big\{ L_{q}^\dagger(k)L_{q}(k),\varrho_t(k)\big\}.
\end{equation}
Defining the jump channels with feedback as $\mathcal{J}_q(k)\rho \equiv L_q(k)\,\rho\,L_q^\dagger(k)$, Eq.~\eqref{eq: jump FB dyn} can be rewritten as
\begin{align}
    \partial_t \varrho_t(k) &= \mathcal{L}_0(k)\,\varrho_t(k) + \mathcal{J}_k(k)\,\varrho_t(k) \nonumber\\
    &\quad + \sum_{q\neq k} \mathcal{J}_k(q)\,\varrho_t(q),
    \label{eq: jump dyn fb with jump channels}
\end{align}
where the sum $\sum_{q\neq k}$ runs over all jump channels $q\in\Sigma$ except $q=k$.

Therefore, Eq.~\eqref{eq: jump dyn fb with jump channels} shows that the rate 
$\partial_t \varrho_t(k)$ has two effective contributions. 
The first term corresponds to no-jump detections, described by $\mathcal{L}_0(k)$; 
these are events in which the last detected jump was $k$ and the subsequent detection 
is a no-jump event. 
The second term corresponds to events in which the last detected transition was $k$ 
and another jump of the same type $k$ is detected, given by 
$\mathcal{J}_k(k)\varrho_t(k) = L_k(k)\,\varrho_t(k)\,L_k^\dagger(k)$. 
Together, these first two terms describe transitions $k \to k$ in the memory space.
The last term corresponds to the event in which the last detected transition was 
$q \neq k$ and a jump of type $k$ is subsequently detected, as described by 
$\mathcal{J}_k(q)\,\varrho_t(q) = L_k(q)\,\varrho_t(q)\,L_k^\dagger(q)$. 
Hence, this term accounts for transitions $q \to k$ in the memory space.

We can connect this general situation with the example provided above. 
For the two-level system coupled to a thermal bath, there are two possible 
transitions: emission or absorption. 
Equation~\eqref{eq: simple example 1} describes the absorption events, whereas 
Eq.~\eqref{eq: simple example -1} corresponds to the emission events.
Let us analyze the absorptions described by Eq.~\eqref{eq: simple example 1} (the same analysis applies to emissions). 
The first term, $-i \big[H_0+\lambda \sigma_x,\varrho_t(1)\big]- \frac{1}{2}   \big\{ L,\varrho_t(1)\big\}$, corresponds to the following event: the last transition was an absorption, and no subsequent jump is detected. 
The second term contains two possibilities: either the last detected transition was an absorption and we detect another absorption, which is described by $L_+\,\varrho_t(1)\,L_+^\dagger$, or the last detected transition was an emission and we then detect an absorption, described by $L_+\,\varrho_t(-1)\,L_+^\dagger$.

We now analyze the time evolution of the probability $P_t(k) = \tr[\varrho_t(k)]$ that the last transition at time $t$ corresponds to $k$. Taking the trace on both sides of Eq.~\eqref{eq: jump FB dyn}, we obtain the following classical master equation~\cite{PhysRevA.107.062206}:
\begin{equation}
\label{eq: evolution of the memory distribution}
\partial_t P_t(k) = \sum_{q \neq k} T_t(k,q) P_t(q) - \sum_{q \neq k} T_t(q,k) P_t(k)~,
\end{equation}
where $\rho_t(k) \equiv \varrho_t(k)/\tr[\varrho_t(k)]$ is the state of the system conditioned on the memory realization $k_t = k$~[Eq.~\eqref{eq: memory-conditioned states - def}], and $T_t(k,q) \equiv \tr\!\left[\mathcal{J}_k(q)\rho_t(q)\right] = \tr\!\left[L_k(q)\rho_t(q)L_k^\dagger(q)\right]$ is the transition rate at which the jump memory $k_t$ changes from $q$ to $k$.
The quantity $\partial_t P_t(k)$ defines a rate equation with two contributions: the first term accounts for trajectories in which the previous transition was $q \neq k$ and the newly detected jump is of type $k$, while the second term describes trajectories in which the last transition was $k$ and the next detected jump is $q \neq k$.

Equation~\eqref{eq: jump FB dyn} describes a general jump-based feedback scheme that can modify both the system Hamiltonian and the jump operators depending on the last detected transition encoded by $k_t$. 
However, as discussed in the example above, one may consider a feedback scheme that modifies only the system Hamiltonian. 
In this case, the jump operators are not affected by the feedback, and we can replace $L_q(k) \to L_q$ for any jump $q \in \Sigma$. 
For this particular case, Eq.~\eqref{eq: jump FB dyn} becomes
\begin{eqnarray}
    \partial_t \varrho_t(k) &=& -i \big[H(k),\varrho_t(k)\big] - \frac{1}{2}  \left\{ \left(\sum_{q \in \Sigma}L_{q}^\dagger L_{q}\right),\varrho_t(k)\right\} \nonumber \\
                             & &  +\ L_k^{}\left(\sum_{q \in \Sigma} \varrho_t(q)\right) L_k^\dagger.
\end{eqnarray}
Finally, as an important consistency check, one should recover the Lindblad master equation~\eqref{QME} in the absence of feedback. 
Indeed, removing the feedback dependence amounts to replacing both $H(k) \to H$ and $L_q(k) \to L_q$ in Eq.~\eqref{eq: jump FB dyn}, corresponding to the situation in which neither the Hamiltonian nor the jump operators are modified by the detected jumps. 
In this case, summing over $k$ on both sides of Eq.~\eqref{eq: jump FB dyn} and using $\sum_{k\in\Sigma} \varrho_t(k) = \bar{\rho}_t$, one straightforwardly recovers Eq.~\eqref{QME}.

\subsection{Previous works on feedback based on jump detections}
\label{subsec: previous jump fb}
A previous work by Wiseman~\cite{Wiseman1994_Feedback} introduced a feedback protocol based on quantum--jump detections: whenever a jump is detected, an instantaneous quantum channel is applied. 
In this framework, the system is continuously monitored via quantum jumps, with the measurement outcome denoted by $x_t$, where $x_t = 0$ corresponds to no-jump events and $x_t = k$ indicates that a jump of type $k$ was detected at time $t$.
Hence, the feedback action consists of applying a quantum channel (a completely positive trace-preserving map) $\mathcal{F}(x_t)$, defined by
\begin{equation}
    \mathcal{F}(x_t) =
    \begin{cases}
        I_d, & \text{if } x_t = 0, \\[4pt]
        e^{\mathcal{K}(k)}, & \text{if } x_t = k ,
    \end{cases}
\end{equation}
where $\mathcal{K}(k)$ is an arbitrary Liouvillian super-operator and $I_d$ is the identity map ($I_d \rho = \rho$).
This feedback scheme can be interpreted as a particular case of the general formalism developed in Ref.~\cite{Rosal2025DeterministicFeedback}, as shown in its Supplemental Material. In this situation, one can derive a closed master equation for the unconditional state of the system, which evolves according to
\begin{eqnarray}
\label{eq: single jump feedback equation}
    \partial_t \bar{\rho}_t &=& -i [H,\bar{\rho}_t]\nonumber\\
    &+& \sum_{k \in \Sigma} \left( e^{\mathcal{K}(k)} \left[L_k \bar{\rho}_tL_k^\dagger\right] - \frac{1}{2}\{L_k^\dagger L_k,\bar{\rho}_t\} \right).
\end{eqnarray}

This approach has several limitations.  
First, the protocol does not retain information about past jumps: the feedback action depends solely on the current detection outcome $x_t$, and no memory of the previous events is preserved. Moreover, the feedback is implemented as an instantaneous quantum channel applied immediately after a detected jump. 
As a consequence, simple protocols such as the one described in Sec.~\ref{subsec:form_simple_example} cannot be captured within this framework, since in that case the detection of an absorption event must be stored and the drive kept active until an emission is detected.  
Second, in quantum-jump monitoring it is far more likely to observe no-jump outcomes ($x_t = 0$) than actual jumps ($x_t = k$)~\cite{Tutorial}. Thus, for most times one has $x_t = 0$, meaning that the feedback is almost never applied.

Therefore, the jump-based feedback described by Eq.~\eqref{eq: jump FB dyn} and the protocols of Eq.~\eqref{eq: single jump feedback equation} are fundamentally different.  
The former implements a memory-based feedback, recording the last transition in a memory $k_t$ and using this information in the feedback action. This protocol discards trivial no-jump events and relies only on the relevant information of the last detected transition.  
In contrast, Eq.~\eqref{eq: single jump feedback equation} describes a memory-less feedback that depends solely on the current detection $x_t$. Consequently, one can write a closed master equation for the unconditional state: without memory of previous jumps, the feedback dynamics can be fully described by the evolution of the system's unconditional state alone.

\section{Full counting statistics for a jump-based feedback dynamics}
\label{sec:results}

\subsection{Hybrid representation of a feedback dynamics}

As shown in Ref.~\cite{Rosal2025MemoryStatFB}, the feedback dynamics can be equivalently described as a hybrid classical–quantum bipartite system. 
In this description, the quantum system subject to quantum-jump detection is represented by the Hilbert space $\mathcal{H}_s$, while the memory $k_t$, which records the last detected jump, is represented by a classical Hilbert space $\mathcal{H}_{\text{cl}}$ with orthonormal basis $\{\ket{k}\}_{k\in\Sigma}$.
We define the \emph{memory-conditioned} states $\rho_t(k)$ and the classical states $\rho_t^{\mathrm{cl}}$ as
\begin{eqnarray}
\label{eq: memory-conditioned states - def}
    \rho_t(k) &\equiv& \frac{\varrho_t(k)}{\tr[\varrho_t(k)]}~,\\
    \rho_t^{\text{cl}}&\equiv& \sum_{k\in\Sigma} P_t(k) \ket{k}\bra{k}~,
\end{eqnarray}
where $P_t(k) = \tr[\varrho_t(k)]$ is the probability distribution of the jump memory $k_t$.

The memory-conditioned states represent the state of the quantum system conditioned on the memory realization $k_t = k$. In fact, since $\bar{\rho}_t = \sum_{k\in\Sigma} \varrho_t(k)$, one has $ \bar{\rho}_t = \sum_{k\in\Sigma} P_t(k) \rho_t(k)$.
Hence, the unconditional state can be seen as an ensemble $\{P_t(k),\rho_t(k)\}$ of the memory-conditioned states. 
Conversely, the classical state $\rho_t^\mathrm{cl}$ represents the state of the memory $k_t$. It corresponds to a diagonal state in the memory basis $\{\ket{k}\}$, where the diagonal elements give the probability of the realization $k_t = k$.

The hybrid system is then described by a tensor Hilbert space $\mathcal{H}_s \otimes \mathcal{H}_{\text{cl}}$, and its bipartite density matrix $\rho_{\text{sm}}(t)$ is defined as 
\begin{equation}
\label{eq: composed quantum-classical state}
    \rho_{\text{sm}}(t) \equiv \sum_{k\in\Sigma} P_t(k) ~\rho_t(k) \otimes \ket{k}\bra{k}~.
\end{equation}
Note that the system's state can be recovered by tracing out the memory space, and analogously, the memory's state can be obtained by tracing out the system space,
\begin{eqnarray}
    \tr_{\mathcal{H}_{\text{cl}}}[\rho_{\text{sm}}(t)] &=&  \sum_{k\in\Sigma} \varrho_t(k) = \bar{\rho}_t~, \\
    \tr_{\mathcal{H}_s}[\rho_{\text{sm}}(t)] &=& \sum_{k\in\Sigma} P_t(k)\, \ket{k}\bra{k} = \rho_t^{\text{cl}}~.
\end{eqnarray}
The advantage of describing the feedback dynamics through the composite bipartite state $\rho_{\text{sm}}(t)$ is that it constitutes a genuine density matrix representing the evolution of a composite system. 
This allows one to employ informational measures to quantify correlations between the system and the memory, as explored in Ref.~\cite{Rosal2025MemoryStatFB}. 
In contrast, the memory-resolved state $\varrho_t(k)$ does not correspond to a density matrix; in particular, it is not normalized, as shown in Eq.~\eqref{eq: memory distribution from SRR state}.

The memory $k_t$ is introduced as a function of the detected measurement record and can be represented as a classical stochastic process with probability distribution $P_t(k)$, or equivalently by the classical state $\rho_t^{\text{cl}}$.
At the same time, the memory also admits a direct physical interpretation. 
Experimentally, the memory is stored in a physical device, and the state $\rho_t^{\text{cl}}$ can therefore be interpreted as the state of this classical system.

The goal of this work is to describe the full statistics of the counting observable $N(t)$ under a general jump-based feedback protocol, as given by Eq.~\eqref{eq: jump FB dyn}, which governs the joint evolution of the system state and the jump statistics. 
To this end, we derive the evolution equation for the composite state $\rho_{\text{sm}}(t)$ and establish its connection to full counting statistics.

\subsection{Lindblad evolution of the composite feedback representation}
\label{sec:results_FB_Master_Equation}

The jump-based feedback master equation~\eqref{eq: jump FB dyn} defines a system of time-local differential equations for the jump-memory states $\varrho_t(k)$, where each derivative $\partial_t \varrho_t(k)$ depends only on the instantaneous states $\varrho_t(q)$ for $q \in \Sigma$.
In contrast, as discussed in Sec.~\ref{subsec: interpretation of the jump-based ME}, one cannot, in general, obtain a closed time-local equation for the unconditional system state $\bar{\rho}_t$. 
In particular, the system is not described by a quantum Markovian Master Equation of the form~\eqref{QME}, since its evolution is coupled to that of the jump memory $k_t$, which must be treated explicitly. As a result, the system exhibits non-Markovian dynamics: the types of past jumps directly influence its subsequent evolution.
Nevertheless, the combined system–memory state $\rho_{\text{sm}}(t)$ obeys a Markovian Master Equation, as stated in the following result:
\begin{result}[Dynamical evolution of the joint state $\rho_{\text{sm}}(t)$]
\label{result1: Joint fb dynamics}
    Given a jump-based feedback protocol as described by Eq.~\eqref{eq: jump FB dyn}, the bipartite state $\rho_{\text{sm}}(t)$ evolves according to the following Markovian Master Equation
    \begin{eqnarray}
        \label{eq: FB dyn in joint system}
        \partial_t\rho_{\text{sm}}(t) &\equiv& \mathbb{L}\rho_{\text{sm}}(t)\nonumber\\
        &=&  -i[\mathbb{H},\rho_{\text{sm}}(t)]+ \sum_{k,q\in\Sigma} \mathcal{D}[\mathbb{L}_{k,q}] \rho_{\text{sm}}(t),
    \end{eqnarray}
    with both the extended Hamiltonian $\mathbb{H}$ and the jump operators $\mathbb{L}_{k,q}$ acting on the joint Hilbert space $\mathcal{H}_s \otimes \mathcal{H}_{\mathrm{cl}}$, defined as
    \begin{eqnarray}
    \label{eq: extended FB Hamiltonian}
        \mathbb{H} &\equiv& \sum_{k\in\Sigma} H(k) \otimes \ket{k}\bra{k}~,\\
    \label{eq: extended FB jump ops}
       \mathbb{L}_{k,q} &\equiv& L_k(q)\otimes \ket{k}\bra{q}~.
    \end{eqnarray}
\end{result}

Equation~\eqref{eq: FB dyn in joint system} is obtained by differentiating both sides of Eq.~\eqref{eq: composed quantum-classical state} with respect to time, then using Eq.~\eqref{eq: jump FB dyn} to evaluate $\partial_t \varrho_t(k)$. Finally, one identifies $\mathbb{H}$ and $\mathbb{L}_{k,q}$ according to Eqs.~\eqref{eq: extended FB Hamiltonian} and \eqref{eq: extended FB jump ops}. For the detailed derivation, see Appendix~\ref{ap: proof of the joint FB dyn}.
Result~\eqref{result1: Joint fb dynamics} shows that the feedback dynamics can be equivalently described either by a Markovian master equation for the joint state $\rho_{\text{sm}}(t)$ with the extended Hamiltonian $\mathbb{H}$ and jump operators $\mathbb{L}_{k,q}$, or by a set of coupled differential equations for the memory-resolved states $\varrho_t(k)$, as given in Eq.~\eqref{eq: jump FB dyn}. 
By solving either of these equations, one can recover the system state as well as the probability $P_t(k)$ of $k_t$.
Result~\eqref{result1: Joint fb dynamics} thus provides a single equation that equivalently captures the same feedback dynamics, at the cost of enlarging the Hilbert space to $\mathcal{H}_s \otimes \mathcal{H}_\text{cl}$.

The extended jump operator $\mathbb{L}_{k,q} = L_k(q) \otimes \ket{k}\bra{q}$ has a clear interpretation in terms of its action on each subsystem. Given that the last detected jump was $q$, the jump operator in the quantum system becomes $L_k(q)$ for any jump $k \in \Sigma$, representing the transition $q \rightarrow k$ in the quantum part. Meanwhile, the corresponding operator in the classical subsystem is $\ket{k}\bra{q}$, representing the memory update, where $k_t$ changes from $q$ to $k$.

Note that $\dim(\mathcal{H}_{\text{cl}})=\#\Sigma$, where $\#\Sigma$ denotes the number of possible jumps, thus directly quantifying the memory cost of the feedback protocol.
The memory-less feedback protocol mentioned in Sec.~\ref{subsec: previous jump fb} has the dynamics described solely in the Hilbert space of the quantum system [Eq.~\eqref{eq: single jump feedback equation}].
On the other hand, the feedback dynamics based on the jump memory $k_t$ is still described by a Markovian equation [Result~\ref{result1: Joint fb dynamics}], but now on the enlarged space $\mathcal{H}_{s}\otimes\mathcal{H}_{\text{cl}}$.

In the next section, we show how to use Result~\eqref{result1: Joint fb dynamics} to compute the full statistics of the stochastic charge $N(t)$ [Eq.~\eqref{eq: stoc charge int}] under a general jump-based feedback. 

\subsection{Connection between the feedback dynamics with FCS}
\label{sec:results_FCS_with_FB}
The main advantage of the hybrid representation of the feedback dynamics described by Eq.~\eqref{eq: FB dyn in joint system} is that we have a one-to-one map between the jumps in the quantum system and jumps in the composite classical-quantum system. In other words, we have a jump $\bar{\rho}_t \rightarrow L_k(q)\bar{\rho}_t L_k^\dagger(q)$ in the quantum system if, and only if, the jump $\mathbb{L}_{k,q}\rho_{\text{sm}}(t)\mathbb{L}_{k,q}^\dagger$ happened in the joint system. 
Consequently, we can unravel the master equation \eqref{eq: FB dyn in joint system} similarly to Eq.~\eqref{QME} in Sec.~\ref{sec:form_FCS}, and it defines the jumps in the extended space $\mathcal{H}_s\otimes \mathcal{H}_{\text{cl}}$.

The extended jump operators $\mathbb{L}_{k,q}$ define the jump channels $\mathbb{J}_{k,q} (\rho_{\text{sm}}) \equiv \mathbb{L}_{k,q}\rho_{\text{sm}}\mathbb{L}_{k,q}^\dagger$ for any density matrix $\rho_{\text{sm}}$ in the space $\mathcal{H}_s\otimes \mathcal{H}_{\text{cl}}$. 
We can define $\tilde{N}_{kq}(t)$ as the total (random) number of jumps in the channel $\mathbb{J}_{k,q}$ over the interval $[0,t]$. 
Therefore, one can define the \emph{extended stochastic charge} as
\begin{equation}
\label{eq: extended stochastic charge}
    \tilde{N}(t) = \sum_{k,q \in \Sigma} \tilde{\nu}_{kq} \tilde{N}_{kq}(t)~.
\end{equation}

Note that Eq.~\eqref{eq: extended stochastic charge} allows us to assign weights not only for the transitions in the quantum system, but also attribute such weights to transitions of the memory $k_t$. 
In particular, if we consider $\tilde{\nu}_{kq} \equiv \nu_k$, where $\nu_k$ are the weights of the system's stochastic charge $N(t)$ [Eq.~\eqref{eq: stoc charge int}], then Eq.~\eqref{eq: extended stochastic charge} becomes
\begin{equation}
    \tilde{N}(t) = \sum_{k\in \Sigma} \nu_{k} \sum_{q\in\Sigma} \tilde{N}_{kq}(t) = \sum_{k\in\Sigma}\nu_k N_k(t) = N(t)~,
\end{equation}
where $\sum_{q\in\Sigma} \tilde{N}_{kq}(t) = N_k(t)$ is the number of jumps in the channel $k$.
It proofs our second main result: 
\begin{result}
\label{result2:FCS with feedback}
With weights $\tilde{\nu}_{kq} \equiv \nu_k$, the extended stochastic charge [Eq.~\eqref{eq: extended stochastic charge}] coincides with that of the quantum subsystem [Eq.~\eqref{eq: stoc charge int}], i.e., $\tilde{N}(t) = N(t)$.
\end{result}

In summary, we have shown that a general feedback protocol based on the last detected jump, initially described by Eq.~\eqref{eq: jump FB dyn}, can be reformulated in terms of an extended system–memory composite state $\rho_{\text{sm}}(t)$, which evolves according to the Markovian Master Equation~\eqref{eq: FB dyn in joint system}.
In this composite system, we introduce a new stochastic charge
$\tilde{N}(t) = \sum_{k,q \in \Sigma} \tilde{\nu}_{kq} \tilde{N}_{kq}(t)$,
whose statistical properties can be characterized using the tools of full counting statistics (FCS) theory~\cite{Tutorial}, as briefly outlined below.
Finally, if we set the weights $\tilde{\nu}_{kq} = \nu_k$, where $\nu_k$ are the weights of the stochastic charge $N(t)$ of the quantum system [Eq.~\eqref{eq: stoc charge int}], then $\tilde{N}(t) = N(t)$, and the statistics of the extended stochastic charge coincide with those of the system's stochastic charge.

Let us review some previous results from the FCS theory \cite{Tutorial} and show how they can be used in the context of a general jump-based feedback protocol. The fluctuations around the expected value $\braket{N(t)}$ of the charge $N(t)$ are captured by the variance $\text{Var}(N(t))$. On the other hand, the rates of the variance and expected value also provide important information about the stochastic charge. We define the average current and noise as 
 \begin{eqnarray}
 \label{eq: def of average current}
       J_t&\equiv&\frac{d}{dt}E[N(t)]~,\\
 \label{eq: def of noise}
       D_t &\equiv& \frac{d}{dt}\text{Var}(N(t))~.
 \end{eqnarray}
One can introduce the super-operator $\tilde{\mathcal{J}}$ acting on the joint space $\mathcal{H}_s \otimes \mathcal{H}_\text{cl}$ as
\begin{equation}
\label{eq: current operator}
    \tilde{\mathcal{J}}\rho_{\text{sm}}(t) \equiv \sum_{k\in\Sigma} \nu_k \sum_{q\in \Sigma}\mathbb{L}_{k,q}\rho_{\text{sm}}(t) \mathbb{L}_{k,q}^\dagger~,
\end{equation}
and the average current of the stochastic charge $N(t)$ under the feedback dynamics can be written as
\begin{equation}
\label{eq: average current}
    J_t = \tr[\tilde{\mathcal{J}}\rho_{\text{sm}}(t)].
\end{equation}

Let us define the stochastic current as
\begin{equation}
    I_t \equiv \frac{dN(t)}{dt}~,
\end{equation}
and consequently one has $J_t = E[I_t]$. 
The time correlations of the stochastic current are captured by the two-point correlation function $F(t,t+\tau)$, that is introduced as
\begin{eqnarray}
    F(t,t+\tau) &\equiv& E[\delta I_t \delta I_{t+\tau}]\\
    & = & E[I_t I_{t+\tau}] - J_t J_{t+\tau}~,\nonumber
\end{eqnarray}
where $\delta I_t \equiv I_t - J_t$ is the current fluctuation. Since $I_t$ is a classical random variable, one has $F(t,t+\tau) = F(t+\tau,t)$, then it is sufficient to consider only $\tau>0$. 
Furthermore, let us define the super-operator $\tilde{\mathcal{H}}$ that acts on $\mathcal{H}_s\otimes\mathcal{H}_\text{cl}$ as
\begin{equation}
\label{eq: kinetic operator}
    \tilde{\mathcal{H}}\rho_{\text{sm}}(t) \equiv \sum_{k\in\Sigma} \nu_k^2 \sum_{q\in \Sigma}\mathbb{L}_{k,q}\rho_{\text{sm}}(t) \mathbb{L}_{k,q}^\dagger~,
\end{equation}
and we introduce $K_t\equiv \tr[\tilde{\mathcal{H}}\rho_{\text{sm}}(t)]$. 

Considering that both system's Hamiltonian and jump operators are time-independent, then the two-point correlation function can be written as 
\begin{equation}
\label{eq: def two-point corr func}
    F(t,t+\tau) = \delta(\tau) K_t + \tr[\tilde{\mathcal{J}} e^{\tau\mathbb{L}}\tilde{\mathcal{J}}\rho_{\text{sm}}(t)] - J_t J_{t+\tau}~,
\end{equation}
where $\mathbb{L}$ is defined in Eq.~\eqref{eq: FB dyn in joint system}, $\tilde{\mathcal{J}}$ and $\tilde{\mathcal{H}}$ are defined in Eqs.~\eqref{eq: current operator} and \eqref{eq: kinetic operator}, respectively. 
The power spectrum, denoted by $S(\omega)$, is defined as the Fourier transform of the two-point correlation function in the steady-state regime, $F(\tau) \equiv \lim_{t \to \infty} F(t,t+\tau)$. Thus,
\begin{eqnarray}
\label{eq: power spectral in the steady state}
S(\omega) &=& \int_{-\infty}^{\infty} e^{-i\omega\tau} F(\tau)\, d\tau \\
&=& K + \int_{-\infty}^{\infty} e^{-i\omega\tau} 
\left( \tr\!\left[\tilde{\mathcal{J}} e^{|\tau|\mathbb{L}}\tilde{\mathcal{J}}\rho_{\text{sm}}^{\text{ss}}\right] - J^2 \right) d\tau~,\nonumber
\end{eqnarray}
where $K \equiv \lim_{t \to \infty} K_t$, and $\rho_{\text{sm}}^{\text{ss}}$ denotes the steady-state of the composite system, satisfying $\mathbb{L}\rho_{\text{sm}}^{\text{ss}} = 0$. One can further show that the zero-frequency component of the power spectrum yields the steady-state noise,
\begin{eqnarray}
\label{eq: noise in the steady state}
D &\equiv& \lim_{t\rightarrow\infty} D_t = S(0) \\
&=& K + 2 \int_{0}^{\infty} 
\left( \tr\!\left[\tilde{\mathcal{J}} e^{\tau\mathbb{L}}\tilde{\mathcal{J}}\rho_{\text{sm}}^{\text{ss}}\right] - J^2 \right) d\tau~.\nonumber
\end{eqnarray}

The steady-state noise defined in Eq.~\eqref{eq: noise in the steady state} quantifies the fluctuations of the stochastic charge $N(t)$ in the long-time regime.
Since the right-hand side of Eq.~\eqref{eq: noise in the steady state} is time-independent, Eq.~\eqref{eq: def of noise} implies that $\text{Var}(N(t)) = D t$ in this regime.
Knowledge of $D$ is essential when designing feedback protocols, as it indicates how fluctuations may hinder or facilitate the ability to steer the system toward desired states or to perform specific tasks.

\section{Application: three-level maser with jump-based feedback }
\label{sec:applicatio_maser}
We now apply the framework developed above to a three-level maser driven by an external field~\cite{PhysRevLett.2.262,PhysRevE.104.L012103}.
We begin by briefly reviewing the different thermodynamic cycles of a quantum thermal machine~\cite{TimeResolved}. We then introduce the dynamics of the three-level maser in the absence of feedback, and subsequently describe a feedback protocol based on its quantum-jump detections. Finally, we discuss how the same feedback strategy can be implemented within the classical maser model.
Our goal is to employ feedback control to select between different thermodynamic regimes of the maser, specifically its operation as an engine or as a refrigerator.
\begin{figure}
    \centering
    \includegraphics[scale=0.62]{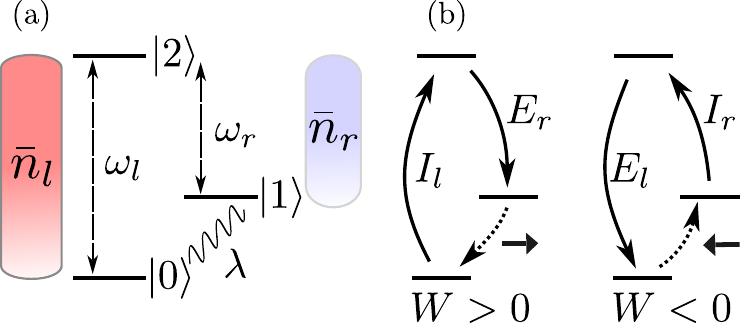}
    \caption{
    { (a)} Three-level maser without feedback. The external drive is continuously applied, inducing Rabi oscillations between states $\ket{0}$ and $\ket{1}$ with strength $\lambda$. The system exchanges thermal photons with the left ($l$) and right ($r$) baths, coupled with rates $\gamma_{\alpha}$ ($\alpha = l, r$).
    { (b)} Thermodynamic cycles of the three-level maser. 
    An engine cycle occurs when the system undergoes the sequence of jumps $\ket{0}\!\rightarrow\!\ket{2}\!\rightarrow\!\ket{1}$ and subsequently evolves from $\ket{1}$ to $\ket{0}$ under the drive, during which work is performed on the external drive ($W > 0$).
    Conversely, a refrigeration cycle corresponds to the reverse trajectory ($W<0$). }
    \label{fig: maser diagram}
\end{figure}

\subsection{Thermodynamic cycles}

A quantum thermal machine consists of a system coupled to two thermal reservoirs, from which it extracts or absorbs energy in the form of work. 
Ref.~\cite{TimeResolved} introduced a framework that resolves its dynamics into thermodynamic cycles, classified as enginelike, coolinglike, or idle.
The thermal machine undergoes either energy injections, corresponding to the absorption of energy from a bath, or energy emissions, corresponding to the release of energy into a bath.
An enginelike cycle corresponds to a process in which the system absorbs energy from one bath, performs work, and discards the remaining energy into the second bath.
Conversely, a coolinglike cycle describes the opposite situation, where work is invested to extract energy from one bath and release it into the other.
Idle cycles correspond to sequences of transitions that do not produce work, such as an emission followed by an absorption from the same bath.

For instance, let us consider a three-level maser as described in Fig.~\ref{fig: maser diagram}(a).
The system's Hilbert space is spanned by the orthonormal basis $\{\ket{0}, \ket{1},\ket{2}\}$. The states $\ket{0}$ and $\ket{2}$, with frequency gap $\omega_l$, are coupled with a thermal bath of Bose-Einstein distribution $\bar{n}_l = (e^{\omega_l/T_l}-1)^{-1}$, and the strength coupling $\gamma_l$. 
Similarly, the states $\ket{1}$ and $\ket{2}$ have a frequency gap $\omega_r$ and are coupled with a thermal bath characterized by $\bar{n}_r$ and $\gamma_r$. 
Here, $T_\alpha$ is the temperature of each bath, with $\alpha = l,r$.
Hence, an \emph{injection} $I_l$ occurs when the system transitions from $\ket{0}$ to $\ket{2}$ by absorbing a thermal photon from the left bath, while an \emph{emission} $E_l$ corresponds to the decay $\ket{2}\to\ket{0}$ accompanied by photon emission into the same bath.
Analogously, the transitions involving states $\ket{1}$ and $\ket{2}$ define emissions $E_r$ and injections $I_r$ associated with the right bath.
Furthermore, the system is subject to an external drive coupling the states $\ket{0}$ and $\ket{1}$, enabling Rabi oscillations between these levels.

When the sequence $I_l E_r$ is observed, the system evolves from $\ket{1}$ to $\ket{0}$ while performing work on the external drive, corresponding to an engine cycle  (see Fig.~\ref{fig: maser diagram}(b)).
Conversely, after the sequence $I_r E_l$, the drive performs work on the system, which evolves from $\ket{0}$ to $\ket{1}$, thereby realizing a cooling cycle.
In our convention, $W>0$ corresponds to work performed by the maser on the external drive, i.e., operating as an engine, whereas $W<0$ corresponds to work done by the external drive on the maser, i.e., operating as a refrigerator.

In what follows, feedback can be used to turn the drive on or off conditioned on the last jump transition, thereby selecting only engine or cooling cycles. 
We first present the maser without feedback, where the drive remains continuously on, and then introduce a feedback protocol based on the jump detections.

\subsection{Maser without feedback}
\label{app_sec:maser_without_FB}
\subsubsection{Dynamics}

Assuming the maser is weakly coupled to the two baths, its evolution (without feedback) is governed by the master equation
\begin{equation}
\label{eq: quantum maser withot fb}
    \partial_t \rho_t = -i[H,\rho_t]+\sum_{k \in\{E_l,I_l,E_r,I_r\}} \mathcal{D}[L_k]\rho_t~,
\end{equation}
where $\rho_t$ is the state of the three-level system. The jump operators of the left bath are given by 
\begin{equation}
    \label{eq: jump op of left bath}
    L_{E_l} = \sqrt{\gamma_l (\bar{n}_l+1)} \ket{0}\bra{2},~ L_{I_l} = \sqrt{\gamma_l \bar{n}_l} \ket{2}\bra{0},
\end{equation}
and for the right bath one has
\begin{equation}
    \label{eq: jump op of right bath}
    L_{E_r} = \sqrt{\gamma_r (\bar{n}_r+1)} \ket{1}\bra{2},~ L_{I_r} = \sqrt{\gamma_r \bar{n}_r} \ket{2}\bra{1}.
\end{equation}
Hence, $L_{I_\alpha}$ and $L_{E_\alpha}$ describe, respectively, injection and emission events associated with bath $\alpha$, where $\alpha = r, l$. 

The system Hamiltonian is $H = H_0 + V $, where $H_0 = (\omega_l - \omega_r)\ket{1}\bra{1} + \omega_l\ket{2}\bra{2}$ is the free Hamiltonian, and $V = 2\lambda \cos{(\omega_d t)} \big(\ket{0}\bra{1} + \ket{1}\bra{0} \big)$ is the interaction due to the external drive with frequency $\omega_d$ and strength $\lambda$. 
In what follows, we work in the interaction picture, effectively removing the free Hamiltonian $H_0$. 
In addition, we move from the interaction picture to a rotating frame at frequency $\Delta \equiv \omega_d -(\omega_l-\omega_r)$, and apply the rotating-wave approximation. In this representation, the system Hamiltonian becomes
\begin{equation}
\label{eq: feedback maser Hamiltonian 2}
    H  = \frac{\Delta}{2}\big(\ket{0}\bra{0}-\ket{1}\bra{1}\big)+ \lambda \big(\ket{0}\bra{1} + \ket{1}\bra{0} \big).
\end{equation}

Equation~\eqref{eq: quantum maser withot fb} describes the three-level maser under a coherent drive, represented by the interaction term $V$ in the system Hamiltonian. One can also consider an incoherent drive, which induces incoherent transitions between the states $\ket{0}$ and $\ket{1}$. This scenario is captured by the following classical reference system~\cite{PhysRevE.104.L012103} with state $\sigma_t$ that evolves according to 
\begin{eqnarray}
\label{eq: classical maser withot fb}
    \partial_t \sigma_t &=& \gamma_c(\mathcal{D}[\ket{0}\bra{1}]\sigma_t+ \mathcal{D}[\ket{1}\bra{0}]\sigma_t) \\
    &&+ \sum_{k \in\{E_l,I_l,E_r,I_r\}} \mathcal{D}[L_k]\sigma_t,\nonumber
\end{eqnarray}
with transition rate $\gamma_c$ given by
\begin{equation}
\label{eq: classical transition rate of the maser}
    \gamma_c \equiv \frac{2 \lambda^2 \Gamma}{\Delta^2 + \Gamma^2}~,
\end{equation}
where $\Gamma \equiv (\gamma_l \bar{n}_l + \gamma_r \bar{n}_r)/2$ is the net decoherence rate. 
This classical system is such that both states $\rho_t$ and $\sigma_t$ have the same populations, but $\sigma_t$ is a diagonal state. Since the current is given by the trace of a super-operator acting on the density matrix [Eq.~\eqref{eq: average current}], then it depends only on the populations of the state. Hence, both classical and quantum systems have the same average current.

\subsubsection{FCS of the three-level maser without feedback}

The thermodynamic properties of the three-level maser have recently attracted considerable attention, both theoretically and experimentally, since it can operate as either an engine or a refrigerator~\cite{TimeResolved}, exhibit violations of thermodynamic uncertainty relations~\cite{PhysRevE.104.L012103}, and serve as a limiting case of a three-body refrigerator system~\cite{Aamir2025}.
By appropriately choosing the weights $\nu_k$ in Eq.~\eqref{eq: stoc charge int}, one can analyze the thermodynamic behavior of the three-level maser. 
In particular, the work $W_t$ performed by the system on the external drive up to time $t$ is defined as
\begin{equation}
\label{eq: stochastic work of the three level maser}
    W_t \equiv \omega_l (N_i^l(t) - N_e^l(t)) - \omega_r (N_e^r(t) - N_i^r(t))~,
\end{equation}
where $N_i^\alpha(t)$ and $N_e^\alpha(t)$ are the (random) number of injections and emissions, respectively, from the bath $\alpha \in \{l,r\}$ in the interval $[0,t]$, and are related to the jump operators $L_{I_\alpha}$ and $L_{E_\alpha}$. 

In other words, the stochastic work corresponds to the difference between the net energy absorbed from or emitted to each bath by the system.
In this case, the average current $J_t = \partial_t \mathbb{E}[W_t]$ represents the instantaneous power delivered to the external drive, $P_t \equiv J_t$.
If $J_t>0$, the maser delivers work to the external drive, thereby operating as an engine. Conversely, if $J_t<0$, the drive performs work on the maser, corresponding to a refrigeration cycle.

One can show (see Appendix~\ref{ap: expressions maser}) that this current is proportional to the difference of the Bose-Einstein distributions of the baths in the steady-state, 
$J_\text{ss}\equiv\lim_{t\to\infty} J_t = \xi (\bar{n}_l - \bar{n}_r)$, 
where $\xi>0$ is a constant depending on the couplings $\gamma_{\alpha}$, the external drive strength $\lambda$, and positive linear combinations of $\bar{n}_\alpha$ with $\alpha \in \{l,r\}$. 
In this sense, the sign of the current is determined by the difference $\bar{n}_l - \bar{n}_r$.
Consequently, if $\omega_l/T_l > \omega_r/T_r$, then $\bar{n}_l>\bar{n}_r$, and heat flows from the left to the right bath ($J>0$), whereas $\omega_l/T_l < \omega_r/T_r$ results in $J<0$.
Our goal is to implement a feedback protocol that selects only engine cycles (a similar protocol can target refrigeration cycles), achieving $J>0$ even when $\bar{n}_l<\bar{n}_r$. 
We then apply the framework developed in Sec.~\ref{sec:results} to compute the average power, noise, and correlations of the stochastic work [Eq.~\eqref{eq: stochastic work of the three level maser}] performed by the maser under a jump-based feedback protocol.

\subsection{Maser with feedback}
\label{app_sec:maser_with_FB}
\subsubsection{Dynamics}

Let us consider the continuous monitoring of quantum jumps in the three-level maser. 
The detection outcome $x_t$ at time $t$ can take five possible values: (i) $x_t = 0$ for a no-jump detection, (ii) $x_t = E_\alpha$ for an emission, and (iii) $x_t = I_\alpha$ for an injection, with $\alpha = l,r$ denoting the left or right bath. 
The jump memory $k_t$ records the last detected transition and can therefore assume four possible values, corresponding to $E_\alpha$ and $I_\alpha$ for $\alpha = l,r$. 
For example, $k_t = E_l$ indicates that, at time $t$, the last jump was an emission to the left-hand bath, $\ket{2}\to\ket{0}$.

We now introduce the following jump-based feedback protocol: if the last jump corresponds to the transition $\ket{2}\!\to\!\ket{1}$ (photon emission to the right bath, corresponding to the event $k_t = E_r$), the external drive between the levels $\ket{0}$ and $\ket{1}$ is turned on; otherwise, the drive is turned off. 
This protocol effectively selects only the engine cycles, excluding the possibility of refrigeration cycles.
Indeed, when a transition $\ket{2}\!\to\!\ket{0}$ ($k_t = E_l$) is detected, the drive is switched off, preventing the completion of a refrigeration cycle. 
Therefore, this feedback protocol leads to the following stochastic Hamiltonian
\begin{equation}
\label{eq: feedback maser Hamiltonian}
    H (k_t) = \frac{\Delta}{2}\big(\ket{0}\bra{0}-\ket{1}\bra{1}\big) + \lambda~ \delta_{k_t,E_r} \big(\ket{0}\bra{1} + \ket{1}\bra{0}\big)~,
\end{equation}
where $\delta_{a,b}$ is the Kronecker delta.
Note that this feedback protocol does not modify the jump operators $L_k$ of the maser [Eq.~\eqref{eq: quantum maser withot fb}], as it solely introduces an additional drive term in the Hamiltonian.
Furthermore, one could select only the refrigeration cycles by turning on the external drive exclusively after detecting a $\ket{2}\!\to\!\ket{0}$ transition and turning it off otherwise.

This feedback dynamics is described by Result~\eqref{result1: Joint fb dynamics} as follows. 
The jump memory is represented by a four-dimensional Hilbert space with an orthonormal basis $\{\ket{E_l},\ket{I_l},\ket{E_r},\ket{I_r}\}$, each state corresponding to a possible realization of $k_t$.
From Eq.~\eqref{eq: extended FB Hamiltonian}, the Hamiltonian of the joint classical-quantum system reads
\begin{eqnarray}
    \mathbb{H} = H_{\text{off}} \otimes \mathcal{P}_\text{off}+H_{\text{on}} \otimes \ket{E_r}\bra{E_r},
\end{eqnarray}
where $\mathcal{P}_\text{off} \equiv \ket{I_l}\bra{I_l} +\ket{E_l}\bra{E_l}+\ket{I_r}\bra{I_r} $ is the projection onto the memory subspace where the drive is turned off, and 
\begin{eqnarray}
    H_{\text{off}} &\equiv& \frac{\Delta}{2}\big(\ket{0}\bra{0}-\ket{1}\bra{1}\big)~,\\
     H_{\text{on}}&\equiv&H_{\text{off}} + \lambda~ \big(\ket{0}\bra{1} + \ket{1}\bra{0}\big)~,
\end{eqnarray}
representing the system Hamiltonian when the drive is off and on, respectively. 
According to Eq.~\eqref{eq: extended FB jump ops}, the jump operators in the joint classical-quantum space take the form $\mathbb{L}_{k,q} = L_k \otimes \ket{k}\bra{q}$, with $k,q \in \{E_l, I_l, E_r, I_r\}$, where $L_k$ is given by Eqs.~\eqref{eq: jump op of left bath} and \eqref{eq: jump op of right bath}. 
The joint state $\rho_{\text{sm}}(t)$ evolves according to Eq.~\eqref{eq: FB dyn in joint system}, fully characterizing the feedback dynamics. 
The unconditional state of the system at time $t$ is obtained by tracing out the classical four-dimensional Hilbert space of the memory, while the probability distribution of the stochastic memory $k_t$ is obtained by tracing out the quantum subsystem. 
The steady-state $\rho_{\text{sm}}^{\text{ss}}$ is defined from Eq.~\eqref{eq: FB dyn in joint system} as the stationary solution satisfying $\mathbb{L}\rho_{\text{sm}}^{\text{ss}} = 0$.

One can implement the same feedback protocol in the classical reference model described by Eq.~\eqref{eq: classical maser withot fb}.
In this case, the feedback action consists of enabling or disabling the incoherent transitions between $\ket{0}$ and $\ket{1}$ depending on the last detected jump. 
Specifically, when the last jump corresponds to the transition $\ket{2}\!\to\!\ket{1}$ ($k_t = E_r$), the jump operators $\sqrt{\gamma_c}\ket{1}\bra{0}$ and $\sqrt{\gamma_c}\ket{0}\bra{1}$ are included to allow the $\ket{0}\leftrightarrow\ket{1}$ transitions; otherwise, these operators are removed.
This action is equivalent to turning an incoherent drive on or off between the states $\ket{0}$ and $\ket{1}$.
This classical scenario also represents a feedback strategy conditioned on the last detected jump and can be described by Result~\eqref{result1: Joint fb dynamics}. Since the classical reference system has a vanishing Hamiltonian [Eq.~\eqref{eq: classical maser withot fb}], Eq.~\eqref{eq: extended FB Hamiltonian} directly implies that $\mathbb{H} = 0$.
For the jump operators $\mathbb{L}_{k,q}$ in the extended space, we consider the same set as in the quantum case, supplemented by the additional operators $\sqrt{\gamma_c}\ket{0}\bra{1}\otimes\ket{E_r}\bra{E_r}$ and $\sqrt{\gamma_c}\ket{1}\bra{0}\otimes\ket{E_r}\bra{E_r}$, which represent the feedback action implemented through the incoherent drive.
As in the dynamics without feedback, this classical feedback scheme yields the same population dynamics as in the quantum-feedback case (and therefore the same average current), while the density matrix remains diagonal at all times.

It is important to remark that monitoring the injection events is required for the stochastic work $W_t$ presented in Eq.~\eqref{eq: stochastic work of the three level maser} to be well defined: the monitoring of injections provides the counting processes $N_l^i(t)$ and $N_r^i(t)$, and the total work corresponds to the energy absorbed from the left bath minus the energy delivered to the right bath.
However, this feedback protocol does not require the detection of injection events; it can in fact be implemented using only emission detections.
Thus, if an emission $\ket{2}\!\to\!\ket{1}$ is detected, the drive is turned on, whereas if an emission $\ket{2}\!\to\!\ket{0}$ is detected, the drive is turned off.
In this case, the continuous monitoring of emissions can be carried out using photon-counting detectors~\cite{PhysRevLett.57.1699,PhysRevLett.57.1696,PhysRevLett.56.2797,PhysRevLett.54.1023}.
\begin{figure}
    \centering
    \includegraphics[scale=0.35]{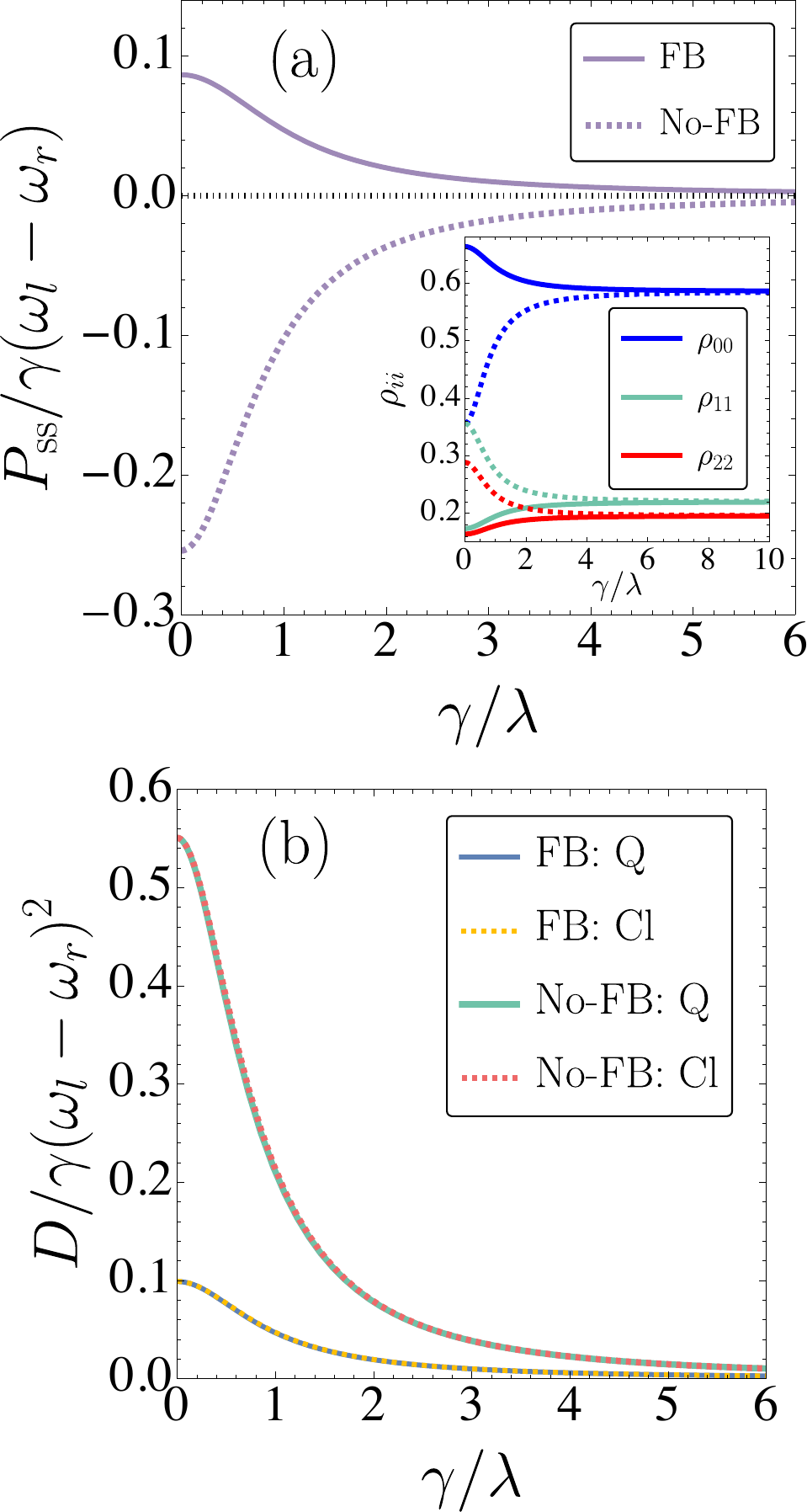}
    \caption{
    {} Selecting only engine cycles in a three-level maser by applying jump-based feedback. 
    We considered $\bar{n}_l = 0.3$, $\bar{n}_r = 8$, $\Delta = 0$, and $\gamma_l = \gamma_r = \gamma$. 
    { (a)}  Average power $P_{\text{ss}} \equiv \lim_{t\to\infty} \partial_t \mathbb{E}[W_t]$ in the steady-state regime. 
    For the maser without feedback, the power is proportional to $\bar{n}_l-\bar{n}_r$ and becomes negative when $\bar{n}_l<\bar{n}_r$ (purple dashed curve). 
    The feedback acts to maintain a positive power even when $\bar{n}_l<\bar{n}_r$ (purple solid curve). 
    The inset shows the steady-state populations of the quantum maser states, where $\rho_{ii} \equiv \bra{i}\rho_{\rm ss}\ket{i}$ is the population of state $\ket{i}$ for $i = 0,1,2$. 
    Solid lines represent populations under feedback dynamics, while dashed lines correspond to the no-feedback case. Colors indicate the same populations in both cases. 
    { (b)} Steady-state noise, $D \equiv \lim_{t\to\infty} \partial_t \mathrm{Var}(W_t)$, of the stochastic work $W_t$ defined in Eq.~\eqref{eq: stochastic work of the three level maser}, shown for both feedback and no-feedback cases with the quantum and classical maser.  }
    \label{fig: diagram and plots}
\end{figure}

\subsubsection{FCS of the three-level maser with feedback}

From now on, we consider a resonant external drive between the states $\ket{0}$ and $\ket{1}$, where $\omega_d = \omega_l - \omega_r$ and thus $\Delta = 0$. 
Hence, $H_{\text{off}} = 0$, and the extended Hamiltonian becomes
\begin{eqnarray}
\mathbb{H} = \lambda~ \big(\ket{0}\bra{1} + \ket{1}\bra{0}\big) \otimes \ket{E_r}\bra{E_r}.
\end{eqnarray}
The feedback steady-state is obtained by solving $\mathbb{L}\rho_{\text{sm}}^{\text{ss}} = 0$, which corresponds to an eigenvector equation, where $\mathbb{L}$ is defined in Eq.~\eqref{eq: FB dyn in joint system}. 
The average steady-state current then follows directly from Eq.~\eqref{eq: average current}.
In this case, the counting observable is the stochastic work $W_t$ [Eq.~\eqref{eq: stochastic work of the three level maser}], so that the current corresponds to the power $P_t$ delivered by the maser to the drive. 
Figure~\ref{fig: diagram and plots}(a) shows the steady-state power for both classical and quantum masers under the feedback protocol described above, compared with the power in the no-feedback case discussed in Sec.~\ref{app_sec:maser_without_FB}.
The classical and quantum masers yield the same steady-state power. 
Under feedback, $P_{\text{ss}}$ is independent of the difference $\bar{n}_l - \bar{n}_r$, resulting in positive power for both $\bar{n}_l \geq \bar{n}_r$ and $\bar{n}_l < \bar{n}_r$ (analytical expressions are given in Appendix~\ref{ap: expressions maser}).

\begin{figure}
    \centering
     \includegraphics[scale=0.35]{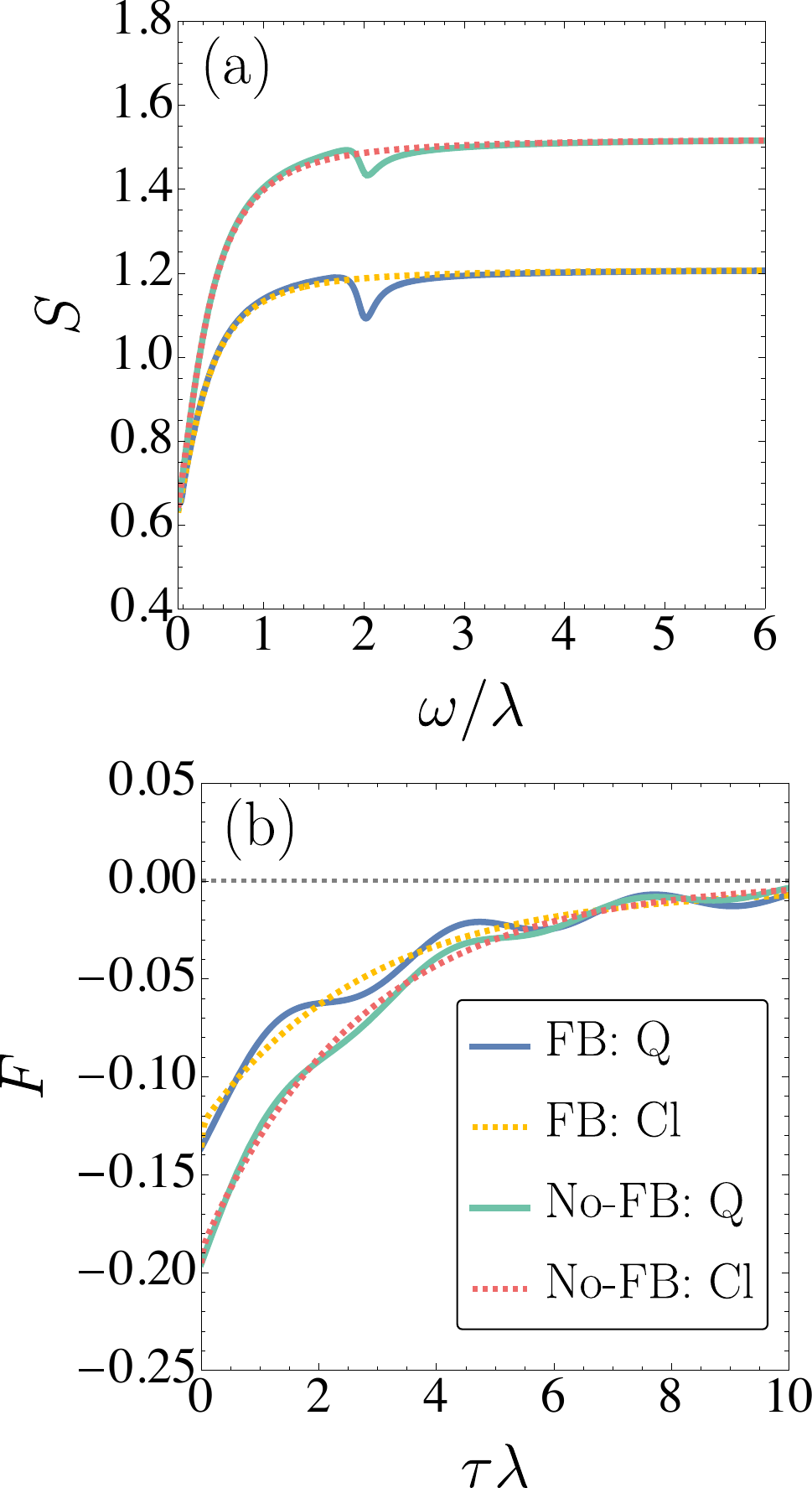}
    \caption{
    { (a)} Power spectrum of the stochastic work defined in Eq.~\eqref{eq: stochastic work of the three level maser}. Remember that the power spectrum is a even function of $\omega$, where $S(\omega) = S(-\omega)$. 
    { (b)} Two-point correlation function in the steady-state defined in Eq.~\eqref{eq: def two-point corr func}.  The parameters used in both (a) and (b) were $\bar{n}_l = 0.3$, $\bar{n}_r = 8$, $\Delta = 0$, $\omega_r = 2 \lambda$, $\omega_l = 4\omega_r$, $\gamma_l = \gamma_r = 0.025 \lambda$, and $\lambda = 1$ set the previous parameters $\gamma,~\omega_r,$ and $\omega_l$ in units of $\lambda$.   }
    \label{fig: spec and curr plots}
\end{figure}

The inset of Fig.~\ref{fig: diagram and plots}(a) shows the feedback steady-state populations of the three quantum-maser levels.
Once we detect the transition $\ket{2} \rightarrow\ket{1}$, then the external drive acts to move the system from $\ket{1}$ to $\ket{0}$ through Rabi oscillations, concluding a engine cycle.
On the other hand, if we detect the transition $\ket{2}\rightarrow\ket{0}$, then there is no external drive to move the system from $\ket{0}$ to $\ket{1}$.
As a consequence, the feedback acts to increase the population of $\ket{0}$ in comparison with the quantum three-level maser without feedback (external drive always on).
Furthermore, note that for $\gamma \gg \lambda$ both the feedback and no-feedback dynamics converge to the same behavior. 
In this regime, the coherent drive is much weaker than the thermal coupling, and both dynamics behave as if the drive were always off.
In contrast, we note that the protocol’s performance is enhanced in the strong-drive regime $\lambda \gg \gamma$, where the power increases as the ratio $\gamma/\lambda$ decreases.

From Eq.~\eqref{eq: noise in the steady state}, one can obtain the noise 
$D \equiv \lim_{t \rightarrow \infty} \partial_t \mathrm{Var}(W_t)$ 
associated with the stochastic charge $W_t$ in the stationary regime, 
as shown in Fig.~\ref{fig: diagram and plots}(b).
Note that the feedback reduces the noise compared to the case without feedback. 
This is because $D$ is related to the fluctuations of the stochastic work $W_t$, which in turn is proportional to the number of possible jump trajectories of the system. 
Since this feedback protocol suppresses all refrigeration cycles, the number of possible trajectories decreases, leading to a reduction in the noise.

\begin{figure}
    \centering
     \includegraphics[width=\linewidth]{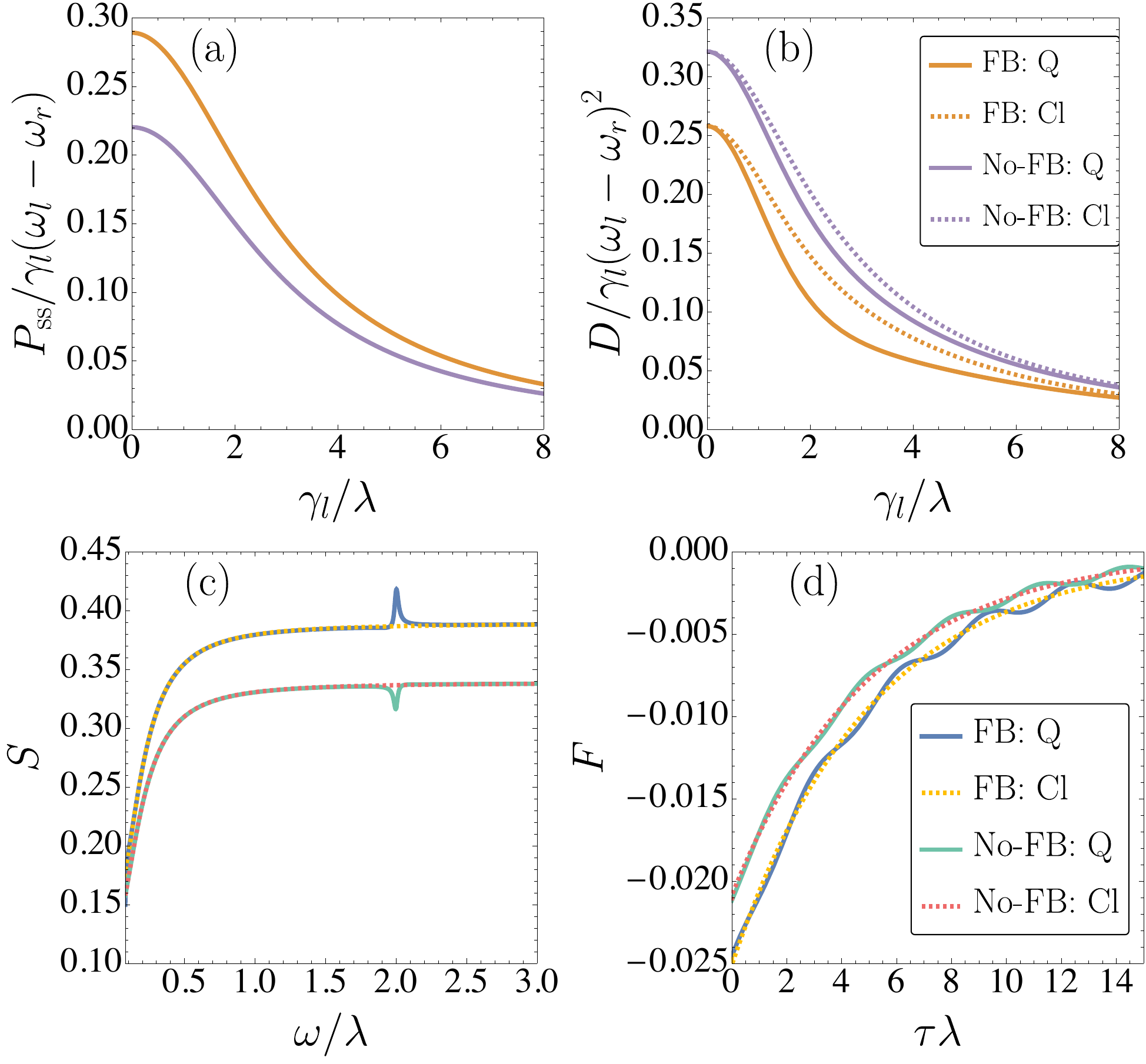}
    \caption{
    {} Enhancing the power of the three-level maser for $\bar{n}_l>\bar{n}_r$.    
    We considered $\bar{n}_l = 0.8$, $\bar{n}_r = 0.1$, $\Delta = 0$, and $ \gamma_r = 5 \gamma_l$. 
    { (a)}  Average power $P_{\text{ss}} \equiv \lim_{t\to\infty} \partial_t \mathbb{E}[W_t]$ in the steady-state regime. 
    { (b)} Steady-state noise, $D \equiv \lim_{t\to\infty} \partial_t \mathrm{Var}(W_t)$.
    { (c)} Power spectrum of the stochastic work, where $S(\omega) = S(-\omega)$. 
    { (d)} Two-point correlation function in the steady-state.  The parameters used in both (c) and (d) were $\omega_r =  \lambda$, $\omega_l = 5 \omega_r$, $\gamma_l = 0.025 \lambda$, and $\lambda = 1$ set the previous parameters in units of $\lambda$.   }    
    
    \label{fig: final example}
\end{figure}

Therefore, the feedback protocol effectively prevents refrigeration cycles, ensuring that the three-level system performs, on average, positive work on the external drive regardless of the relative values of the Bose-Einstein distributions of the baths. 
In other words, the protocol converts the information obtained from jump detection into work extracted by the external drive, thereby enforcing the operation of the maser as an engine.

Equation~\eqref{eq: power spectral in the steady state} provides the power spectrum of the stochastic work defined in Eq.~\eqref{eq: stochastic work of the three level maser}, as shown in Fig.~\ref{fig: spec and curr plots}(a). 
Note that the dips of $S(\omega)$ are associated with the energy gaps in the system Hamiltonian, while the spectral width reflects the coupling strengths to the thermal baths~\cite{Tutorial}. 
In this case, the dips occur at $\omega = \pm 2\lambda$, corresponding to the energy gap of the Hamiltonian defined in Eq.~\eqref{eq: feedback maser Hamiltonian} for $\Delta = 0$.
The constant $K = \tr[\tilde{\mathcal{H}}\rho_{\text{sm}}^{\text{ss}}]$, where $\tilde{\mathcal{H}}$ is defined in Eq.~\eqref{eq: kinetic operator}, determines the background of the power spectrum, corresponding to the white-noise contribution in the two-point correlation function defined in Eq.~\eqref{eq: def two-point corr func}. 
As shown in Fig.~\ref{fig: spec and curr plots}(a), the feedback protocol reduces the background $K$. 
Finally, Fig.~\ref{fig: spec and curr plots}(b) shows the two-point correlation function for the same parameters used in the power spectrum and considering the stationary regime. 
One observes that the stochastic current exhibits negative time correlations ($F(\tau) \leq 0$), reflecting the low probability of detecting another event (either an injection or an emission) immediately after the first one.

As a final example, let us consider the case $\bar{n}_l > \bar{n}_r$, where the three-level maser without feedback already operates, on average, as an engine and therefore exhibits positive power.
The feedback action further suppresses cooling cycles, thereby enhancing the steady-state power of the maser, as shown in Fig.~\ref{fig: final example}(a).
Note that, unlike in Fig.~\ref{fig: diagram and plots}(b), where the noise in the quantum and classical regimes was nearly identical, the noise here exhibits a significant difference between the two descriptions, and the feedback action reduces the noise in this case as well.
Figure~\ref{fig: final example}(c) shows that the feedback in the quantum system produces a valley in the spectral density, in contrast with the dip observed in the no-feedback case and in Fig.~\ref{fig: spec and curr plots}(a). This demonstrates that the feedback modifies the spectral density and, consequently, introduces both correlations and anti-correlations in the jump detections.

\section{Conclusion}
\label{sec:conclusion}

We have shown that a general feedback protocol based on the last detected jump can be dynamically described by a Lindblad master equation in an extended hybrid classical-quantum space (Result~\eqref{result1: Joint fb dynamics}). 
In this framework, the quantum component corresponds to the system under jump monitoring, while the classical component represents the memory that records the last detected jump. 
This result is significant both computationally and conceptually: computationally, it shows that the feedback dynamics can be obtained from a standard Markovian Master Equation; conceptually, it demonstrates that strongly non-Markovian feedback processes can nonetheless be represented as Markovian evolutions in an extended space.
As our second main result, we show that this representation in the hybrid extended space allows a complete characterization of the full counting statistics of any counting observable of a quantum system under general jump-based feedback dynamics (Result~\eqref{result2:FCS with feedback}). This establishes a direct connection between feedback-controlled dynamics and the standard full counting statistics framework for quantum jumps.

The example presented illustrates how the proposed framework enables a complete thermodynamic characterization of systems subject to jump-based feedback dynamics, providing analytical expressions for key physical quantities such as steady-state populations and currents. This application further demonstrates how jump detections can be exploited for work extraction, establishing feedback as a relevant thermodynamic mechanism in thermal machines. In particular, the feedback protocol considered here selectively preserves the engine cycles of the three-level maser while suppressing its refrigeration cycles.

Both Results~\eqref{result1: Joint fb dynamics} and \eqref{result2:FCS with feedback} are particularly relevant at the interface of thermodynamics and quantum information theory, where information gained through measurements can be converted into work via appropriately designed feedback protocols.
This framework provides analytical tools to quantify key system observables under general jump-based feedback dynamics -- features that had remained unexplored due to the absence of a unified formalism capable of describing feedback processes in quantum systems.
Further work has focused on feedback strategies that depend not only on the last jump channel but also on the time elapsed since its detection -- a direction currently under investigation. 

Our formalism remains applicable whenever the dynamics admits a jump-based unraveling, consisting of continuous no-jump evolution interspersed with random discrete jumps, as is the case for weakly coupled systems described by Lindblad master equations [Eq.~\eqref{QME}]. An interesting direction for future work is to extend the formalism to strongly coupled open quantum systems beyond the Lindblad framework.

\begin{acknowledgments}
This research is primarily supported by the U.S. Department of Energy (DOE), Office of Science, Basic Energy Sciences (BES) under Award No. DE-SC0025516. 
P.P.P. acknowledges funding from the Swiss National Science Foundation (Eccellenza Professorial Fellowship PCEFP2\_194268).
\end{acknowledgments}

\bibliography{feedback-refs}

\begin{thebibliography}{52}%
\makeatletter
\providecommand \@ifxundefined [1]{%
 \@ifx{#1\undefined}
}%
\providecommand \@ifnum [1]{%
 \ifnum #1\expandafter \@firstoftwo
 \else \expandafter \@secondoftwo
 \fi
}%
\providecommand \@ifx [1]{%
 \ifx #1\expandafter \@firstoftwo
 \else \expandafter \@secondoftwo
 \fi
}%
\providecommand \natexlab [1]{#1}%
\providecommand \enquote  [1]{``#1''}%
\providecommand \bibnamefont  [1]{#1}%
\providecommand \bibfnamefont [1]{#1}%
\providecommand \citenamefont [1]{#1}%
\providecommand \href@noop [0]{\@secondoftwo}%
\providecommand \href [0]{\begingroup \@sanitize@url \@href}%
\providecommand \@href[1]{\@@startlink{#1}\@@href}%
\providecommand \@@href[1]{\endgroup#1\@@endlink}%
\providecommand \@sanitize@url [0]{\catcode `\\12\catcode `\$12\catcode `\&12\catcode `\#12\catcode `\^12\catcode `\_12\catcode `\%12\relax}%
\providecommand \@@startlink[1]{}%
\providecommand \@@endlink[0]{}%
\providecommand \url  [0]{\begingroup\@sanitize@url \@url }%
\providecommand \@url [1]{\endgroup\@href {#1}{\urlprefix }}%
\providecommand \urlprefix  [0]{URL }%
\providecommand \Eprint [0]{\href }%
\providecommand \doibase [0]{http://dx.doi.org/}%
\providecommand \selectlanguage [0]{\@gobble}%
\providecommand \bibinfo  [0]{\@secondoftwo}%
\providecommand \bibfield  [0]{\@secondoftwo}%
\providecommand \translation [1]{[#1]}%
\providecommand \BibitemOpen [0]{}%
\providecommand \bibitemStop [0]{}%
\providecommand \bibitemNoStop [0]{.\EOS\space}%
\providecommand \EOS [0]{\spacefactor3000\relax}%
\providecommand \BibitemShut  [1]{\csname bibitem#1\endcsname}%
\let\auto@bib@innerbib\@empty
\bibitem [{\citenamefont {Landi}\ \emph {et~al.}(2024)\citenamefont {Landi}, \citenamefont {Kewming}, \citenamefont {Mitchison},\ and\ \citenamefont {Potts}}]{Tutorial}%
  \BibitemOpen
  \bibfield  {author} {\bibinfo {author} {\bibfnamefont {G.~T.}\ \bibnamefont {Landi}}, \bibinfo {author} {\bibfnamefont {M.~J.}\ \bibnamefont {Kewming}}, \bibinfo {author} {\bibfnamefont {M.~T.}\ \bibnamefont {Mitchison}}, \ and\ \bibinfo {author} {\bibfnamefont {P.~P.}\ \bibnamefont {Potts}},\ }\href {\doibase 10.1103/PRXQuantum.5.020201} {\bibfield  {journal} {\bibinfo  {journal} {PRX Quantum}\ }\textbf {\bibinfo {volume} {5}},\ \bibinfo {pages} {020201} (\bibinfo {year} {2024})}\BibitemShut {NoStop}%
\bibitem [{\citenamefont {Goan}\ \emph {et~al.}(2001)\citenamefont {Goan}, \citenamefont {Milburn}, \citenamefont {Wiseman},\ and\ \citenamefont {Bi~Sun}}]{PhysRevB.63.125326}%
  \BibitemOpen
  \bibfield  {author} {\bibinfo {author} {\bibfnamefont {H.-S.}\ \bibnamefont {Goan}}, \bibinfo {author} {\bibfnamefont {G.~J.}\ \bibnamefont {Milburn}}, \bibinfo {author} {\bibfnamefont {H.~M.}\ \bibnamefont {Wiseman}}, \ and\ \bibinfo {author} {\bibfnamefont {H.}~\bibnamefont {Bi~Sun}},\ }\href {\doibase 10.1103/PhysRevB.63.125326} {\bibfield  {journal} {\bibinfo  {journal} {Phys. Rev. B}\ }\textbf {\bibinfo {volume} {63}},\ \bibinfo {pages} {125326} (\bibinfo {year} {2001})}\BibitemShut {NoStop}%
\bibitem [{\citenamefont {Gammelmark}\ and\ \citenamefont {M\o{}lmer}(2014)}]{PhysRevLett.112.170401}%
  \BibitemOpen
  \bibfield  {author} {\bibinfo {author} {\bibfnamefont {S.}~\bibnamefont {Gammelmark}}\ and\ \bibinfo {author} {\bibfnamefont {K.}~\bibnamefont {M\o{}lmer}},\ }\href {\doibase 10.1103/PhysRevLett.112.170401} {\bibfield  {journal} {\bibinfo  {journal} {Phys. Rev. Lett.}\ }\textbf {\bibinfo {volume} {112}},\ \bibinfo {pages} {170401} (\bibinfo {year} {2014})}\BibitemShut {NoStop}%
\bibitem [{\citenamefont {Santos}\ and\ \citenamefont {Landi}(2025)}]{PhysRevA.111.042415}%
  \BibitemOpen
  \bibfield  {author} {\bibinfo {author} {\bibfnamefont {L.~F.}\ \bibnamefont {Santos}}\ and\ \bibinfo {author} {\bibfnamefont {G.~T.}\ \bibnamefont {Landi}},\ }\href {\doibase 10.1103/PhysRevA.111.042415} {\bibfield  {journal} {\bibinfo  {journal} {Phys. Rev. A}\ }\textbf {\bibinfo {volume} {111}},\ \bibinfo {pages} {042415} (\bibinfo {year} {2025})}\BibitemShut {NoStop}%
\bibitem [{\citenamefont {Landi}(2021)}]{PhysRevB.104.195408}%
  \BibitemOpen
  \bibfield  {author} {\bibinfo {author} {\bibfnamefont {G.~T.}\ \bibnamefont {Landi}},\ }\href {\doibase 10.1103/PhysRevB.104.195408} {\bibfield  {journal} {\bibinfo  {journal} {Phys. Rev. B}\ }\textbf {\bibinfo {volume} {104}},\ \bibinfo {pages} {195408} (\bibinfo {year} {2021})}\BibitemShut {NoStop}%
\bibitem [{\citenamefont {Skinner}\ and\ \citenamefont {Dunkel}(2021)}]{PhysRevLett.127.198101}%
  \BibitemOpen
  \bibfield  {author} {\bibinfo {author} {\bibfnamefont {D.~J.}\ \bibnamefont {Skinner}}\ and\ \bibinfo {author} {\bibfnamefont {J.}~\bibnamefont {Dunkel}},\ }\href {\doibase 10.1103/PhysRevLett.127.198101} {\bibfield  {journal} {\bibinfo  {journal} {Phys. Rev. Lett.}\ }\textbf {\bibinfo {volume} {127}},\ \bibinfo {pages} {198101} (\bibinfo {year} {2021})}\BibitemShut {NoStop}%
\bibitem [{\citenamefont {Schulz}\ \emph {et~al.}(2023)\citenamefont {Schulz}, \citenamefont {Chevallier},\ and\ \citenamefont {Albert}}]{PhysRevB.107.245406}%
  \BibitemOpen
  \bibfield  {author} {\bibinfo {author} {\bibfnamefont {F.}~\bibnamefont {Schulz}}, \bibinfo {author} {\bibfnamefont {D.}~\bibnamefont {Chevallier}}, \ and\ \bibinfo {author} {\bibfnamefont {M.}~\bibnamefont {Albert}},\ }\href {\doibase 10.1103/PhysRevB.107.245406} {\bibfield  {journal} {\bibinfo  {journal} {Phys. Rev. B}\ }\textbf {\bibinfo {volume} {107}},\ \bibinfo {pages} {245406} (\bibinfo {year} {2023})}\BibitemShut {NoStop}%
\bibitem [{\citenamefont {Bergquist}\ \emph {et~al.}(1986)\citenamefont {Bergquist}, \citenamefont {Hulet}, \citenamefont {Itano},\ and\ \citenamefont {Wineland}}]{PhysRevLett.57.1699}%
  \BibitemOpen
  \bibfield  {author} {\bibinfo {author} {\bibfnamefont {J.~C.}\ \bibnamefont {Bergquist}}, \bibinfo {author} {\bibfnamefont {R.~G.}\ \bibnamefont {Hulet}}, \bibinfo {author} {\bibfnamefont {W.~M.}\ \bibnamefont {Itano}}, \ and\ \bibinfo {author} {\bibfnamefont {D.~J.}\ \bibnamefont {Wineland}},\ }\href {\doibase 10.1103/PhysRevLett.57.1699} {\bibfield  {journal} {\bibinfo  {journal} {Phys. Rev. Lett.}\ }\textbf {\bibinfo {volume} {57}},\ \bibinfo {pages} {1699} (\bibinfo {year} {1986})}\BibitemShut {NoStop}%
\bibitem [{\citenamefont {Sauter}\ \emph {et~al.}(1986)\citenamefont {Sauter}, \citenamefont {Neuhauser}, \citenamefont {Blatt},\ and\ \citenamefont {Toschek}}]{PhysRevLett.57.1696}%
  \BibitemOpen
  \bibfield  {author} {\bibinfo {author} {\bibfnamefont {T.}~\bibnamefont {Sauter}}, \bibinfo {author} {\bibfnamefont {W.}~\bibnamefont {Neuhauser}}, \bibinfo {author} {\bibfnamefont {R.}~\bibnamefont {Blatt}}, \ and\ \bibinfo {author} {\bibfnamefont {P.~E.}\ \bibnamefont {Toschek}},\ }\href {\doibase 10.1103/PhysRevLett.57.1696} {\bibfield  {journal} {\bibinfo  {journal} {Phys. Rev. Lett.}\ }\textbf {\bibinfo {volume} {57}},\ \bibinfo {pages} {1696} (\bibinfo {year} {1986})}\BibitemShut {NoStop}%
\bibitem [{\citenamefont {Nagourney}\ \emph {et~al.}(1986)\citenamefont {Nagourney}, \citenamefont {Sandberg},\ and\ \citenamefont {Dehmelt}}]{PhysRevLett.56.2797}%
  \BibitemOpen
  \bibfield  {author} {\bibinfo {author} {\bibfnamefont {W.}~\bibnamefont {Nagourney}}, \bibinfo {author} {\bibfnamefont {J.}~\bibnamefont {Sandberg}}, \ and\ \bibinfo {author} {\bibfnamefont {H.}~\bibnamefont {Dehmelt}},\ }\href {\doibase 10.1103/PhysRevLett.56.2797} {\bibfield  {journal} {\bibinfo  {journal} {Phys. Rev. Lett.}\ }\textbf {\bibinfo {volume} {56}},\ \bibinfo {pages} {2797} (\bibinfo {year} {1986})}\BibitemShut {NoStop}%
\bibitem [{\citenamefont {Cook}\ and\ \citenamefont {Kimble}(1985)}]{PhysRevLett.54.1023}%
  \BibitemOpen
  \bibfield  {author} {\bibinfo {author} {\bibfnamefont {R.~J.}\ \bibnamefont {Cook}}\ and\ \bibinfo {author} {\bibfnamefont {H.~J.}\ \bibnamefont {Kimble}},\ }\href {\doibase 10.1103/PhysRevLett.54.1023} {\bibfield  {journal} {\bibinfo  {journal} {Phys. Rev. Lett.}\ }\textbf {\bibinfo {volume} {54}},\ \bibinfo {pages} {1023} (\bibinfo {year} {1985})}\BibitemShut {NoStop}%
\bibitem [{\citenamefont {Rosal}\ \emph {et~al.}(2026)\citenamefont {Rosal}, \citenamefont {Potts},\ and\ \citenamefont {Landi}}]{Rosal2025DeterministicFeedback}%
  \BibitemOpen
  \bibfield  {author} {\bibinfo {author} {\bibfnamefont {A.~J.~B.}\ \bibnamefont {Rosal}}, \bibinfo {author} {\bibfnamefont {P.~P.}\ \bibnamefont {Potts}}, \ and\ \bibinfo {author} {\bibfnamefont {G.~T.}\ \bibnamefont {Landi}},\ }\href {\doibase 10.1103/pg2p-h529} {\bibfield  {journal} {\bibinfo  {journal} {Phys. Rev. Lett.}\ } (\bibinfo {year} {2026}),\ 10.1103/pg2p-h529}\BibitemShut {NoStop}%
\bibitem [{\citenamefont {Wiseman}(1994)}]{Wiseman1994_Feedback}%
  \BibitemOpen
  \bibfield  {author} {\bibinfo {author} {\bibfnamefont {H.~M.}\ \bibnamefont {Wiseman}},\ }\href {\doibase 10.1103/PhysRevA.49.2133} {\bibfield  {journal} {\bibinfo  {journal} {Phys. Rev. A}\ }\textbf {\bibinfo {volume} {49}},\ \bibinfo {pages} {2133} (\bibinfo {year} {1994})}\BibitemShut {NoStop}%
\bibitem [{\citenamefont {Rosal}\ and\ \citenamefont {Landi}(2025)}]{Rosal2025MemoryStatFB}%
  \BibitemOpen
  \bibfield  {author} {\bibinfo {author} {\bibfnamefont {A.~J.~B.}\ \bibnamefont {Rosal}}\ and\ \bibinfo {author} {\bibfnamefont {G.~T.}\ \bibnamefont {Landi}},\ }\href {https://arxiv.org/abs/} {\bibfield  {journal} {\bibinfo  {journal} {arXiv preprint arXiv:}\ } (\bibinfo {year} {2025})},\ \bibinfo {note} {quant-ph}\BibitemShut {NoStop}%
\bibitem [{\citenamefont {D'Urso}\ \emph {et~al.}(2003)\citenamefont {D'Urso}, \citenamefont {Odom},\ and\ \citenamefont {Gabrielse}}]{PhysRevLett.90.043001}%
  \BibitemOpen
  \bibfield  {author} {\bibinfo {author} {\bibfnamefont {B.}~\bibnamefont {D'Urso}}, \bibinfo {author} {\bibfnamefont {B.}~\bibnamefont {Odom}}, \ and\ \bibinfo {author} {\bibfnamefont {G.}~\bibnamefont {Gabrielse}},\ }\href {\doibase 10.1103/PhysRevLett.90.043001} {\bibfield  {journal} {\bibinfo  {journal} {Phys. Rev. Lett.}\ }\textbf {\bibinfo {volume} {90}},\ \bibinfo {pages} {043001} (\bibinfo {year} {2003})}\BibitemShut {NoStop}%
\bibitem [{\citenamefont {Steck}\ \emph {et~al.}(2006)\citenamefont {Steck}, \citenamefont {Jacobs}, \citenamefont {Mabuchi}, \citenamefont {Habib},\ and\ \citenamefont {Bhattacharya}}]{PhysRevA.74.012322}%
  \BibitemOpen
  \bibfield  {author} {\bibinfo {author} {\bibfnamefont {D.~A.}\ \bibnamefont {Steck}}, \bibinfo {author} {\bibfnamefont {K.}~\bibnamefont {Jacobs}}, \bibinfo {author} {\bibfnamefont {H.}~\bibnamefont {Mabuchi}}, \bibinfo {author} {\bibfnamefont {S.}~\bibnamefont {Habib}}, \ and\ \bibinfo {author} {\bibfnamefont {T.}~\bibnamefont {Bhattacharya}},\ }\href {\doibase 10.1103/PhysRevA.74.012322} {\bibfield  {journal} {\bibinfo  {journal} {Phys. Rev. A}\ }\textbf {\bibinfo {volume} {74}},\ \bibinfo {pages} {012322} (\bibinfo {year} {2006})}\BibitemShut {NoStop}%
\bibitem [{\citenamefont {Bushev}\ \emph {et~al.}(2006)\citenamefont {Bushev}, \citenamefont {Rotter}, \citenamefont {Wilson}, \citenamefont {Dubin}, \citenamefont {Becher}, \citenamefont {Eschner}, \citenamefont {Blatt}, \citenamefont {Steixner}, \citenamefont {Rabl},\ and\ \citenamefont {Zoller}}]{PhysRevLett.96.043003}%
  \BibitemOpen
  \bibfield  {author} {\bibinfo {author} {\bibfnamefont {P.}~\bibnamefont {Bushev}}, \bibinfo {author} {\bibfnamefont {D.}~\bibnamefont {Rotter}}, \bibinfo {author} {\bibfnamefont {A.}~\bibnamefont {Wilson}}, \bibinfo {author} {\bibfnamefont {F.~m.~c.}\ \bibnamefont {Dubin}}, \bibinfo {author} {\bibfnamefont {C.}~\bibnamefont {Becher}}, \bibinfo {author} {\bibfnamefont {J.}~\bibnamefont {Eschner}}, \bibinfo {author} {\bibfnamefont {R.}~\bibnamefont {Blatt}}, \bibinfo {author} {\bibfnamefont {V.}~\bibnamefont {Steixner}}, \bibinfo {author} {\bibfnamefont {P.}~\bibnamefont {Rabl}}, \ and\ \bibinfo {author} {\bibfnamefont {P.}~\bibnamefont {Zoller}},\ }\href {\doibase 10.1103/PhysRevLett.96.043003} {\bibfield  {journal} {\bibinfo  {journal} {Phys. Rev. Lett.}\ }\textbf {\bibinfo {volume} {96}},\ \bibinfo {pages} {043003} (\bibinfo {year} {2006})}\BibitemShut {NoStop}%
\bibitem [{\citenamefont {Jacobs}\ \emph {et~al.}(2015)\citenamefont {Jacobs}, \citenamefont {Nurdin}, \citenamefont {Strauch},\ and\ \citenamefont {James}}]{PhysRevA.91.043812}%
  \BibitemOpen
  \bibfield  {author} {\bibinfo {author} {\bibfnamefont {K.}~\bibnamefont {Jacobs}}, \bibinfo {author} {\bibfnamefont {H.~I.}\ \bibnamefont {Nurdin}}, \bibinfo {author} {\bibfnamefont {F.~W.}\ \bibnamefont {Strauch}}, \ and\ \bibinfo {author} {\bibfnamefont {M.}~\bibnamefont {James}},\ }\href {\doibase 10.1103/PhysRevA.91.043812} {\bibfield  {journal} {\bibinfo  {journal} {Phys. Rev. A}\ }\textbf {\bibinfo {volume} {91}},\ \bibinfo {pages} {043812} (\bibinfo {year} {2015})}\BibitemShut {NoStop}%
\bibitem [{\citenamefont {Frimmer}\ \emph {et~al.}(2016)\citenamefont {Frimmer}, \citenamefont {Gieseler},\ and\ \citenamefont {Novotny}}]{PhysRevLett.117.163601}%
  \BibitemOpen
  \bibfield  {author} {\bibinfo {author} {\bibfnamefont {M.}~\bibnamefont {Frimmer}}, \bibinfo {author} {\bibfnamefont {J.}~\bibnamefont {Gieseler}}, \ and\ \bibinfo {author} {\bibfnamefont {L.}~\bibnamefont {Novotny}},\ }\href {\doibase 10.1103/PhysRevLett.117.163601} {\bibfield  {journal} {\bibinfo  {journal} {Phys. Rev. Lett.}\ }\textbf {\bibinfo {volume} {117}},\ \bibinfo {pages} {163601} (\bibinfo {year} {2016})}\BibitemShut {NoStop}%
\bibitem [{\citenamefont {Guo}\ \emph {et~al.}(2019)\citenamefont {Guo}, \citenamefont {Norte},\ and\ \citenamefont {Gr\"oblacher}}]{PhysRevLett.123.223602}%
  \BibitemOpen
  \bibfield  {author} {\bibinfo {author} {\bibfnamefont {J.}~\bibnamefont {Guo}}, \bibinfo {author} {\bibfnamefont {R.}~\bibnamefont {Norte}}, \ and\ \bibinfo {author} {\bibfnamefont {S.}~\bibnamefont {Gr\"oblacher}},\ }\href {\doibase 10.1103/PhysRevLett.123.223602} {\bibfield  {journal} {\bibinfo  {journal} {Phys. Rev. Lett.}\ }\textbf {\bibinfo {volume} {123}},\ \bibinfo {pages} {223602} (\bibinfo {year} {2019})}\BibitemShut {NoStop}%
\bibitem [{\citenamefont {Buffoni}\ \emph {et~al.}(2019)\citenamefont {Buffoni}, \citenamefont {Solfanelli}, \citenamefont {Verrucchi}, \citenamefont {Cuccoli},\ and\ \citenamefont {Campisi}}]{PhysRevLett.122.070603}%
  \BibitemOpen
  \bibfield  {author} {\bibinfo {author} {\bibfnamefont {L.}~\bibnamefont {Buffoni}}, \bibinfo {author} {\bibfnamefont {A.}~\bibnamefont {Solfanelli}}, \bibinfo {author} {\bibfnamefont {P.}~\bibnamefont {Verrucchi}}, \bibinfo {author} {\bibfnamefont {A.}~\bibnamefont {Cuccoli}}, \ and\ \bibinfo {author} {\bibfnamefont {M.}~\bibnamefont {Campisi}},\ }\href {\doibase 10.1103/PhysRevLett.122.070603} {\bibfield  {journal} {\bibinfo  {journal} {Phys. Rev. Lett.}\ }\textbf {\bibinfo {volume} {122}},\ \bibinfo {pages} {070603} (\bibinfo {year} {2019})}\BibitemShut {NoStop}%
\bibitem [{\citenamefont {Manikandan}\ and\ \citenamefont {Qvarfort}(2023)}]{PhysRevA.107.023516}%
  \BibitemOpen
  \bibfield  {author} {\bibinfo {author} {\bibfnamefont {S.~K.}\ \bibnamefont {Manikandan}}\ and\ \bibinfo {author} {\bibfnamefont {S.}~\bibnamefont {Qvarfort}},\ }\href {\doibase 10.1103/PhysRevA.107.023516} {\bibfield  {journal} {\bibinfo  {journal} {Phys. Rev. A}\ }\textbf {\bibinfo {volume} {107}},\ \bibinfo {pages} {023516} (\bibinfo {year} {2023})}\BibitemShut {NoStop}%
\bibitem [{\citenamefont {De~Sousa}\ \emph {et~al.}(2025)\citenamefont {De~Sousa}, \citenamefont {Bakhshinezhad}, \citenamefont {Annby-Andersson}, \citenamefont {Samuelsson}, \citenamefont {Potts},\ and\ \citenamefont {Jarzynski}}]{CoolingGui}%
  \BibitemOpen
  \bibfield  {author} {\bibinfo {author} {\bibfnamefont {G.}~\bibnamefont {De~Sousa}}, \bibinfo {author} {\bibfnamefont {P.}~\bibnamefont {Bakhshinezhad}}, \bibinfo {author} {\bibfnamefont {B.}~\bibnamefont {Annby-Andersson}}, \bibinfo {author} {\bibfnamefont {P.}~\bibnamefont {Samuelsson}}, \bibinfo {author} {\bibfnamefont {P.~P.}\ \bibnamefont {Potts}}, \ and\ \bibinfo {author} {\bibfnamefont {C.}~\bibnamefont {Jarzynski}},\ }\href {\doibase 10.1103/PhysRevE.111.014152} {\bibfield  {journal} {\bibinfo  {journal} {Phys. Rev. E}\ }\textbf {\bibinfo {volume} {111}},\ \bibinfo {pages} {014152} (\bibinfo {year} {2025})}\BibitemShut {NoStop}%
\bibitem [{\citenamefont {Hopkins}\ \emph {et~al.}(2003)\citenamefont {Hopkins}, \citenamefont {Jacobs}, \citenamefont {Habib},\ and\ \citenamefont {Schwab}}]{PhysRevB.68.235328}%
  \BibitemOpen
  \bibfield  {author} {\bibinfo {author} {\bibfnamefont {A.}~\bibnamefont {Hopkins}}, \bibinfo {author} {\bibfnamefont {K.}~\bibnamefont {Jacobs}}, \bibinfo {author} {\bibfnamefont {S.}~\bibnamefont {Habib}}, \ and\ \bibinfo {author} {\bibfnamefont {K.}~\bibnamefont {Schwab}},\ }\href {\doibase 10.1103/PhysRevB.68.235328} {\bibfield  {journal} {\bibinfo  {journal} {Phys. Rev. B}\ }\textbf {\bibinfo {volume} {68}},\ \bibinfo {pages} {235328} (\bibinfo {year} {2003})}\BibitemShut {NoStop}%
\bibitem [{\citenamefont {Rossi}\ \emph {et~al.}(2017)\citenamefont {Rossi}, \citenamefont {Kralj}, \citenamefont {Zippilli}, \citenamefont {Natali}, \citenamefont {Borrielli}, \citenamefont {Pandraud}, \citenamefont {Serra}, \citenamefont {Di~Giuseppe},\ and\ \citenamefont {Vitali}}]{PhysRevLett.119.123603}%
  \BibitemOpen
  \bibfield  {author} {\bibinfo {author} {\bibfnamefont {M.}~\bibnamefont {Rossi}}, \bibinfo {author} {\bibfnamefont {N.}~\bibnamefont {Kralj}}, \bibinfo {author} {\bibfnamefont {S.}~\bibnamefont {Zippilli}}, \bibinfo {author} {\bibfnamefont {R.}~\bibnamefont {Natali}}, \bibinfo {author} {\bibfnamefont {A.}~\bibnamefont {Borrielli}}, \bibinfo {author} {\bibfnamefont {G.}~\bibnamefont {Pandraud}}, \bibinfo {author} {\bibfnamefont {E.}~\bibnamefont {Serra}}, \bibinfo {author} {\bibfnamefont {G.}~\bibnamefont {Di~Giuseppe}}, \ and\ \bibinfo {author} {\bibfnamefont {D.}~\bibnamefont {Vitali}},\ }\href {\doibase 10.1103/PhysRevLett.119.123603} {\bibfield  {journal} {\bibinfo  {journal} {Phys. Rev. Lett.}\ }\textbf {\bibinfo {volume} {119}},\ \bibinfo {pages} {123603} (\bibinfo {year} {2017})}\BibitemShut {NoStop}%
\bibitem [{\citenamefont {Zippilli}\ \emph {et~al.}(2018)\citenamefont {Zippilli}, \citenamefont {Kralj}, \citenamefont {Rossi}, \citenamefont {Di~Giuseppe},\ and\ \citenamefont {Vitali}}]{PhysRevA.98.023828}%
  \BibitemOpen
  \bibfield  {author} {\bibinfo {author} {\bibfnamefont {S.}~\bibnamefont {Zippilli}}, \bibinfo {author} {\bibfnamefont {N.}~\bibnamefont {Kralj}}, \bibinfo {author} {\bibfnamefont {M.}~\bibnamefont {Rossi}}, \bibinfo {author} {\bibfnamefont {G.}~\bibnamefont {Di~Giuseppe}}, \ and\ \bibinfo {author} {\bibfnamefont {D.}~\bibnamefont {Vitali}},\ }\href {\doibase 10.1103/PhysRevA.98.023828} {\bibfield  {journal} {\bibinfo  {journal} {Phys. Rev. A}\ }\textbf {\bibinfo {volume} {98}},\ \bibinfo {pages} {023828} (\bibinfo {year} {2018})}\BibitemShut {NoStop}%
\bibitem [{\citenamefont {Sarovar}\ \emph {et~al.}(2004)\citenamefont {Sarovar}, \citenamefont {Ahn}, \citenamefont {Jacobs},\ and\ \citenamefont {Milburn}}]{QECwithFeedback}%
  \BibitemOpen
  \bibfield  {author} {\bibinfo {author} {\bibfnamefont {M.}~\bibnamefont {Sarovar}}, \bibinfo {author} {\bibfnamefont {C.}~\bibnamefont {Ahn}}, \bibinfo {author} {\bibfnamefont {K.}~\bibnamefont {Jacobs}}, \ and\ \bibinfo {author} {\bibfnamefont {G.~J.}\ \bibnamefont {Milburn}},\ }\href {\doibase 10.1103/PhysRevA.69.052324} {\bibfield  {journal} {\bibinfo  {journal} {Phys. Rev. A}\ }\textbf {\bibinfo {volume} {69}},\ \bibinfo {pages} {052324} (\bibinfo {year} {2004})}\BibitemShut {NoStop}%
\bibitem [{\citenamefont {Marcos}\ \emph {et~al.}(2020)\citenamefont {Marcos}, \citenamefont {Smith}, \citenamefont {Bednorz},\ and\ \citenamefont {Yunger~Halpern}}]{QuantumThermodynamicsForQuantumComputing}%
  \BibitemOpen
  \bibfield  {author} {\bibinfo {author} {\bibfnamefont {D.}~\bibnamefont {Marcos}}, \bibinfo {author} {\bibfnamefont {A.}~\bibnamefont {Smith}}, \bibinfo {author} {\bibfnamefont {A.}~\bibnamefont {Bednorz}}, \ and\ \bibinfo {author} {\bibfnamefont {N.}~\bibnamefont {Yunger~Halpern}},\ }\href {\doibase 10.1103/PhysRevX.10.041013} {\bibfield  {journal} {\bibinfo  {journal} {Phys. Rev. X}\ }\textbf {\bibinfo {volume} {10}},\ \bibinfo {pages} {041013} (\bibinfo {year} {2020})}\BibitemShut {NoStop}%
\bibitem [{\citenamefont {Vijay}\ \emph {et~al.}(2012)\citenamefont {Vijay}, \citenamefont {Macklin}, \citenamefont {Slichter}, \citenamefont {Weber}, \citenamefont {Murch}, \citenamefont {Naik}, \citenamefont {Korotkov},\ and\ \citenamefont {Siddiqi}}]{Vijay2012}%
  \BibitemOpen
  \bibfield  {author} {\bibinfo {author} {\bibfnamefont {R.}~\bibnamefont {Vijay}}, \bibinfo {author} {\bibfnamefont {C.}~\bibnamefont {Macklin}}, \bibinfo {author} {\bibfnamefont {D.~H.}\ \bibnamefont {Slichter}}, \bibinfo {author} {\bibfnamefont {S.~J.}\ \bibnamefont {Weber}}, \bibinfo {author} {\bibfnamefont {K.~W.}\ \bibnamefont {Murch}}, \bibinfo {author} {\bibfnamefont {R.}~\bibnamefont {Naik}}, \bibinfo {author} {\bibfnamefont {A.~N.}\ \bibnamefont {Korotkov}}, \ and\ \bibinfo {author} {\bibfnamefont {I.}~\bibnamefont {Siddiqi}},\ }\href {\doibase 10.1038/nature11505} {\bibfield  {journal} {\bibinfo  {journal} {Nature}\ }\textbf {\bibinfo {volume} {490}},\ \bibinfo {pages} {77} (\bibinfo {year} {2012})}\BibitemShut {NoStop}%
\bibitem [{\citenamefont {Rist\`e}\ \emph {et~al.}(2012)\citenamefont {Rist\`e}, \citenamefont {Bultink}, \citenamefont {Lehnert},\ and\ \citenamefont {DiCarlo}}]{DiscreteFeedback2}%
  \BibitemOpen
  \bibfield  {author} {\bibinfo {author} {\bibfnamefont {D.}~\bibnamefont {Rist\`e}}, \bibinfo {author} {\bibfnamefont {C.~C.}\ \bibnamefont {Bultink}}, \bibinfo {author} {\bibfnamefont {K.~W.}\ \bibnamefont {Lehnert}}, \ and\ \bibinfo {author} {\bibfnamefont {L.}~\bibnamefont {DiCarlo}},\ }\href {\doibase 10.1103/PhysRevLett.109.240502} {\bibfield  {journal} {\bibinfo  {journal} {Phys. Rev. Lett.}\ }\textbf {\bibinfo {volume} {109}},\ \bibinfo {pages} {240502} (\bibinfo {year} {2012})}\BibitemShut {NoStop}%
\bibitem [{\citenamefont {Zippilli}\ \emph {et~al.}(2003)\citenamefont {Zippilli}, \citenamefont {Vitali}, \citenamefont {Tombesi},\ and\ \citenamefont {Raimond}}]{PhysRevA.67.052101}%
  \BibitemOpen
  \bibfield  {author} {\bibinfo {author} {\bibfnamefont {S.}~\bibnamefont {Zippilli}}, \bibinfo {author} {\bibfnamefont {D.}~\bibnamefont {Vitali}}, \bibinfo {author} {\bibfnamefont {P.}~\bibnamefont {Tombesi}}, \ and\ \bibinfo {author} {\bibfnamefont {J.-M.}\ \bibnamefont {Raimond}},\ }\href {\doibase 10.1103/PhysRevA.67.052101} {\bibfield  {journal} {\bibinfo  {journal} {Phys. Rev. A}\ }\textbf {\bibinfo {volume} {67}},\ \bibinfo {pages} {052101} (\bibinfo {year} {2003})}\BibitemShut {NoStop}%
\bibitem [{\citenamefont {Rossi}\ \emph {et~al.}(2018)\citenamefont {Rossi}, \citenamefont {Kralj}, \citenamefont {Zippilli}, \citenamefont {Natali}, \citenamefont {Borrielli}, \citenamefont {Pandraud}, \citenamefont {Serra}, \citenamefont {Di~Giuseppe},\ and\ \citenamefont {Vitali}}]{PhysRevLett.120.073601}%
  \BibitemOpen
  \bibfield  {author} {\bibinfo {author} {\bibfnamefont {M.}~\bibnamefont {Rossi}}, \bibinfo {author} {\bibfnamefont {N.}~\bibnamefont {Kralj}}, \bibinfo {author} {\bibfnamefont {S.}~\bibnamefont {Zippilli}}, \bibinfo {author} {\bibfnamefont {R.}~\bibnamefont {Natali}}, \bibinfo {author} {\bibfnamefont {A.}~\bibnamefont {Borrielli}}, \bibinfo {author} {\bibfnamefont {G.}~\bibnamefont {Pandraud}}, \bibinfo {author} {\bibfnamefont {E.}~\bibnamefont {Serra}}, \bibinfo {author} {\bibfnamefont {G.}~\bibnamefont {Di~Giuseppe}}, \ and\ \bibinfo {author} {\bibfnamefont {D.}~\bibnamefont {Vitali}},\ }\href {\doibase 10.1103/PhysRevLett.120.073601} {\bibfield  {journal} {\bibinfo  {journal} {Phys. Rev. Lett.}\ }\textbf {\bibinfo {volume} {120}},\ \bibinfo {pages} {073601} (\bibinfo {year} {2018})}\BibitemShut {NoStop}%
\bibitem [{\citenamefont {Vidrighin}\ \emph {et~al.}(2016)\citenamefont {Vidrighin}, \citenamefont {Dahlsten}, \citenamefont {Barbieri}, \citenamefont {Kim}, \citenamefont {Vedral},\ and\ \citenamefont {Walmsley}}]{QuantumDemon1}%
  \BibitemOpen
  \bibfield  {author} {\bibinfo {author} {\bibfnamefont {M.~D.}\ \bibnamefont {Vidrighin}}, \bibinfo {author} {\bibfnamefont {O.}~\bibnamefont {Dahlsten}}, \bibinfo {author} {\bibfnamefont {M.}~\bibnamefont {Barbieri}}, \bibinfo {author} {\bibfnamefont {M.~S.}\ \bibnamefont {Kim}}, \bibinfo {author} {\bibfnamefont {V.}~\bibnamefont {Vedral}}, \ and\ \bibinfo {author} {\bibfnamefont {I.~A.}\ \bibnamefont {Walmsley}},\ }\href {\doibase 10.1103/PhysRevLett.116.050401} {\bibfield  {journal} {\bibinfo  {journal} {Phys. Rev. Lett.}\ }\textbf {\bibinfo {volume} {116}},\ \bibinfo {pages} {050401} (\bibinfo {year} {2016})}\BibitemShut {NoStop}%
\bibitem [{\citenamefont {Naghiloo}\ \emph {et~al.}(2018)\citenamefont {Naghiloo}, \citenamefont {Alonso}, \citenamefont {Romito}, \citenamefont {Lutz},\ and\ \citenamefont {Murch}}]{QuantumDemon2}%
  \BibitemOpen
  \bibfield  {author} {\bibinfo {author} {\bibfnamefont {M.}~\bibnamefont {Naghiloo}}, \bibinfo {author} {\bibfnamefont {J.~J.}\ \bibnamefont {Alonso}}, \bibinfo {author} {\bibfnamefont {A.}~\bibnamefont {Romito}}, \bibinfo {author} {\bibfnamefont {E.}~\bibnamefont {Lutz}}, \ and\ \bibinfo {author} {\bibfnamefont {K.~W.}\ \bibnamefont {Murch}},\ }\href {\doibase 10.1103/PhysRevLett.121.030604} {\bibfield  {journal} {\bibinfo  {journal} {Phys. Rev. Lett.}\ }\textbf {\bibinfo {volume} {121}},\ \bibinfo {pages} {030604} (\bibinfo {year} {2018})}\BibitemShut {NoStop}%
\bibitem [{\citenamefont {Ribezzi-Crivellari}\ and\ \citenamefont {Ritort}(2019)}]{QuantumDemon3}%
  \BibitemOpen
  \bibfield  {author} {\bibinfo {author} {\bibfnamefont {M.}~\bibnamefont {Ribezzi-Crivellari}}\ and\ \bibinfo {author} {\bibfnamefont {F.}~\bibnamefont {Ritort}},\ }\href {\doibase 10.1038/s41567-019-0481-0} {\bibfield  {journal} {\bibinfo  {journal} {Nature Physics}\ }\textbf {\bibinfo {volume} {15}},\ \bibinfo {pages} {660} (\bibinfo {year} {2019})}\BibitemShut {NoStop}%
\bibitem [{\citenamefont {Sagawa}\ and\ \citenamefont {Ueda}(2010)}]{SecondLawFeedback1}%
  \BibitemOpen
  \bibfield  {author} {\bibinfo {author} {\bibfnamefont {T.}~\bibnamefont {Sagawa}}\ and\ \bibinfo {author} {\bibfnamefont {M.}~\bibnamefont {Ueda}},\ }\href {\doibase 10.1103/PhysRevLett.104.090602} {\bibfield  {journal} {\bibinfo  {journal} {Phys. Rev. Lett.}\ }\textbf {\bibinfo {volume} {104}},\ \bibinfo {pages} {090602} (\bibinfo {year} {2010})}\BibitemShut {NoStop}%
\bibitem [{\citenamefont {Sagawa}\ and\ \citenamefont {Ueda}(2012)}]{SecondLawFeedback2}%
  \BibitemOpen
  \bibfield  {author} {\bibinfo {author} {\bibfnamefont {T.}~\bibnamefont {Sagawa}}\ and\ \bibinfo {author} {\bibfnamefont {M.}~\bibnamefont {Ueda}},\ }\href {\doibase 10.1103/PhysRevLett.109.180602} {\bibfield  {journal} {\bibinfo  {journal} {Phys. Rev. Lett.}\ }\textbf {\bibinfo {volume} {109}},\ \bibinfo {pages} {180602} (\bibinfo {year} {2012})}\BibitemShut {NoStop}%
\bibitem [{\citenamefont {Funo}\ \emph {et~al.}(2013)\citenamefont {Funo}, \citenamefont {Watanabe},\ and\ \citenamefont {Ueda}}]{SecondLawFeedback3}%
  \BibitemOpen
  \bibfield  {author} {\bibinfo {author} {\bibfnamefont {K.}~\bibnamefont {Funo}}, \bibinfo {author} {\bibfnamefont {Y.}~\bibnamefont {Watanabe}}, \ and\ \bibinfo {author} {\bibfnamefont {M.}~\bibnamefont {Ueda}},\ }\href {\doibase 10.1103/PhysRevE.88.052121} {\bibfield  {journal} {\bibinfo  {journal} {Phys. Rev. E}\ }\textbf {\bibinfo {volume} {88}},\ \bibinfo {pages} {052121} (\bibinfo {year} {2013})}\BibitemShut {NoStop}%
\bibitem [{\citenamefont {Minev}\ \emph {et~al.}(2019)\citenamefont {Minev}, \citenamefont {Mundhada}, \citenamefont {Shankar}, \citenamefont {Reinhold}, \citenamefont {Guti\'errez-J\'auregui}, \citenamefont {Schoelkopf}, \citenamefont {Mirrahimi}, \citenamefont {Carmichael},\ and\ \citenamefont {Devoret}}]{Minev2019CatchingReverseQuantumJump}%
  \BibitemOpen
  \bibfield  {author} {\bibinfo {author} {\bibfnamefont {Z.~K.}\ \bibnamefont {Minev}}, \bibinfo {author} {\bibfnamefont {S.~O.}\ \bibnamefont {Mundhada}}, \bibinfo {author} {\bibfnamefont {S.}~\bibnamefont {Shankar}}, \bibinfo {author} {\bibfnamefont {P.}~\bibnamefont {Reinhold}}, \bibinfo {author} {\bibfnamefont {R.}~\bibnamefont {Guti\'errez-J\'auregui}}, \bibinfo {author} {\bibfnamefont {R.~J.}\ \bibnamefont {Schoelkopf}}, \bibinfo {author} {\bibfnamefont {M.}~\bibnamefont {Mirrahimi}}, \bibinfo {author} {\bibfnamefont {H.~J.}\ \bibnamefont {Carmichael}}, \ and\ \bibinfo {author} {\bibfnamefont {M.~H.}\ \bibnamefont {Devoret}},\ }\href {\doibase 10.1038/s41586-019-1287-z} {\bibfield  {journal} {\bibinfo  {journal} {Nature}\ }\textbf {\bibinfo {volume} {570}},\ \bibinfo {pages} {200} (\bibinfo {year} {2019})},\ \Eprint {http://arxiv.org/abs/1902.10355} {arXiv:1902.10355 [quant-ph]} \BibitemShut {NoStop}%
\bibitem [{\citenamefont {Hofmann}\ \emph {et~al.}(2016)\citenamefont {Hofmann}, \citenamefont {Maisi}, \citenamefont {Gold}, \citenamefont {Kr\"ahenmann}, \citenamefont {R\"ossler}, \citenamefont {Basset}, \citenamefont {M\"arki}, \citenamefont {Reichl}, \citenamefont {Wegscheider}, \citenamefont {Ensslin},\ and\ \citenamefont {Ihn}}]{PhysRevLett.117.206803}%
  \BibitemOpen
  \bibfield  {author} {\bibinfo {author} {\bibfnamefont {A.}~\bibnamefont {Hofmann}}, \bibinfo {author} {\bibfnamefont {V.~F.}\ \bibnamefont {Maisi}}, \bibinfo {author} {\bibfnamefont {C.}~\bibnamefont {Gold}}, \bibinfo {author} {\bibfnamefont {T.}~\bibnamefont {Kr\"ahenmann}}, \bibinfo {author} {\bibfnamefont {C.}~\bibnamefont {R\"ossler}}, \bibinfo {author} {\bibfnamefont {J.}~\bibnamefont {Basset}}, \bibinfo {author} {\bibfnamefont {P.}~\bibnamefont {M\"arki}}, \bibinfo {author} {\bibfnamefont {C.}~\bibnamefont {Reichl}}, \bibinfo {author} {\bibfnamefont {W.}~\bibnamefont {Wegscheider}}, \bibinfo {author} {\bibfnamefont {K.}~\bibnamefont {Ensslin}}, \ and\ \bibinfo {author} {\bibfnamefont {T.}~\bibnamefont {Ihn}},\ }\href {\doibase 10.1103/PhysRevLett.117.206803} {\bibfield  {journal} {\bibinfo  {journal} {Phys. Rev. Lett.}\ }\textbf {\bibinfo {volume} {117}},\ \bibinfo {pages} {206803} (\bibinfo {year} {2016})}\BibitemShut {NoStop}%
\bibitem [{\citenamefont {Ye}\ \emph {et~al.}(2025)\citenamefont {Ye}, \citenamefont {Ellaboudy},\ and\ \citenamefont {Nichol}}]{PhysRevApplied.23.044063}%
  \BibitemOpen
  \bibfield  {author} {\bibinfo {author} {\bibfnamefont {F.}~\bibnamefont {Ye}}, \bibinfo {author} {\bibfnamefont {A.}~\bibnamefont {Ellaboudy}}, \ and\ \bibinfo {author} {\bibfnamefont {J.~M.}\ \bibnamefont {Nichol}},\ }\href {\doibase 10.1103/PhysRevApplied.23.044063} {\bibfield  {journal} {\bibinfo  {journal} {Phys. Rev. Appl.}\ }\textbf {\bibinfo {volume} {23}},\ \bibinfo {pages} {044063} (\bibinfo {year} {2025})}\BibitemShut {NoStop}%
\bibitem [{\citenamefont {Kewming}\ \emph {et~al.}(2024)\citenamefont {Kewming}, \citenamefont {Kiely}, \citenamefont {Campbell},\ and\ \citenamefont {Landi}}]{FPTLandi}%
  \BibitemOpen
  \bibfield  {author} {\bibinfo {author} {\bibfnamefont {M.~J.}\ \bibnamefont {Kewming}}, \bibinfo {author} {\bibfnamefont {A.}~\bibnamefont {Kiely}}, \bibinfo {author} {\bibfnamefont {S.}~\bibnamefont {Campbell}}, \ and\ \bibinfo {author} {\bibfnamefont {G.~T.}\ \bibnamefont {Landi}},\ }\href {\doibase 10.1103/PhysRevA.109.L050202} {\bibfield  {journal} {\bibinfo  {journal} {Phys. Rev. A}\ }\textbf {\bibinfo {volume} {109}},\ \bibinfo {pages} {L050202} (\bibinfo {year} {2024})}\BibitemShut {NoStop}%
\bibitem [{\citenamefont {Fiusa}\ \emph {et~al.}(2025{\natexlab{a}})\citenamefont {Fiusa}, \citenamefont {Harunari}, \citenamefont {Hegde},\ and\ \citenamefont {Landi}}]{Fiusa2025CountingObservables}%
  \BibitemOpen
  \bibfield  {author} {\bibinfo {author} {\bibfnamefont {G.}~\bibnamefont {Fiusa}}, \bibinfo {author} {\bibfnamefont {P.~E.}\ \bibnamefont {Harunari}}, \bibinfo {author} {\bibfnamefont {A.~S.}\ \bibnamefont {Hegde}}, \ and\ \bibinfo {author} {\bibfnamefont {G.~T.}\ \bibnamefont {Landi}},\ }\href {https://arxiv.org/abs/2505.06208} {\bibfield  {journal} {\bibinfo  {journal} {arXiv preprint arXiv:2505.06208}\ } (\bibinfo {year} {2025}{\natexlab{a}})},\ \bibinfo {note} {cond-mat.stat-mech}\BibitemShut {NoStop}%
\bibitem [{\citenamefont {Fiusa}\ \emph {et~al.}(2025{\natexlab{b}})\citenamefont {Fiusa}, \citenamefont {Harunari}, \citenamefont {Hegde},\ and\ \citenamefont {Landi}}]{Fiusa2025FrameworkFluctuatingTimes}%
  \BibitemOpen
  \bibfield  {author} {\bibinfo {author} {\bibfnamefont {G.}~\bibnamefont {Fiusa}}, \bibinfo {author} {\bibfnamefont {P.~E.}\ \bibnamefont {Harunari}}, \bibinfo {author} {\bibfnamefont {A.~S.}\ \bibnamefont {Hegde}}, \ and\ \bibinfo {author} {\bibfnamefont {G.~T.}\ \bibnamefont {Landi}},\ }\href {https://arxiv.org/pdf/2506.05160} {\bibfield  {journal} {\bibinfo  {journal} {arXiv preprint arXiv:2506.05160}\ } (\bibinfo {year} {2025}{\natexlab{b}})},\ \bibinfo {note} {cond-mat.stat-mech}\BibitemShut {NoStop}%
\bibitem [{\citenamefont {Prech}\ \emph {et~al.}(2023)\citenamefont {Prech}, \citenamefont {Johansson}, \citenamefont {Nyholm}, \citenamefont {Landi}, \citenamefont {Verdozzi}, \citenamefont {Samuelsson},\ and\ \citenamefont {Potts}}]{PhysRevResearch.5.023155}%
  \BibitemOpen
  \bibfield  {author} {\bibinfo {author} {\bibfnamefont {K.}~\bibnamefont {Prech}}, \bibinfo {author} {\bibfnamefont {P.}~\bibnamefont {Johansson}}, \bibinfo {author} {\bibfnamefont {E.}~\bibnamefont {Nyholm}}, \bibinfo {author} {\bibfnamefont {G.~T.}\ \bibnamefont {Landi}}, \bibinfo {author} {\bibfnamefont {C.}~\bibnamefont {Verdozzi}}, \bibinfo {author} {\bibfnamefont {P.}~\bibnamefont {Samuelsson}}, \ and\ \bibinfo {author} {\bibfnamefont {P.~P.}\ \bibnamefont {Potts}},\ }\href {\doibase 10.1103/PhysRevResearch.5.023155} {\bibfield  {journal} {\bibinfo  {journal} {Phys. Rev. Res.}\ }\textbf {\bibinfo {volume} {5}},\ \bibinfo {pages} {023155} (\bibinfo {year} {2023})}\BibitemShut {NoStop}%
\bibitem [{\citenamefont {Nishiyama}\ and\ \citenamefont {Hasegawa}(2024)}]{PhysRevE.109.044114}%
  \BibitemOpen
  \bibfield  {author} {\bibinfo {author} {\bibfnamefont {T.}~\bibnamefont {Nishiyama}}\ and\ \bibinfo {author} {\bibfnamefont {Y.}~\bibnamefont {Hasegawa}},\ }\href {\doibase 10.1103/PhysRevE.109.044114} {\bibfield  {journal} {\bibinfo  {journal} {Phys. Rev. E}\ }\textbf {\bibinfo {volume} {109}},\ \bibinfo {pages} {044114} (\bibinfo {year} {2024})}\BibitemShut {NoStop}%
\bibitem [{\citenamefont {Sayrin}\ \emph {et~al.}(2011)\citenamefont {Sayrin}, \citenamefont {Dotsenko}, \citenamefont {Zhou}, \citenamefont {Peaudecerf}, \citenamefont {Rybarczyk}, \citenamefont {Gleyzes}, \citenamefont {Rouchon}, \citenamefont {Mirrahimi}, \citenamefont {Amini}, \citenamefont {Brune}, \citenamefont {Raimond},\ and\ \citenamefont {Haroche}}]{Sayrin2011RealTimeFeedback}%
  \BibitemOpen
  \bibfield  {author} {\bibinfo {author} {\bibfnamefont {C.}~\bibnamefont {Sayrin}}, \bibinfo {author} {\bibfnamefont {I.}~\bibnamefont {Dotsenko}}, \bibinfo {author} {\bibfnamefont {X.}~\bibnamefont {Zhou}}, \bibinfo {author} {\bibfnamefont {B.}~\bibnamefont {Peaudecerf}}, \bibinfo {author} {\bibfnamefont {T.}~\bibnamefont {Rybarczyk}}, \bibinfo {author} {\bibfnamefont {S.}~\bibnamefont {Gleyzes}}, \bibinfo {author} {\bibfnamefont {P.}~\bibnamefont {Rouchon}}, \bibinfo {author} {\bibfnamefont {M.}~\bibnamefont {Mirrahimi}}, \bibinfo {author} {\bibfnamefont {H.}~\bibnamefont {Amini}}, \bibinfo {author} {\bibfnamefont {M.}~\bibnamefont {Brune}}, \bibinfo {author} {\bibfnamefont {J.-M.}\ \bibnamefont {Raimond}}, \ and\ \bibinfo {author} {\bibfnamefont {S.}~\bibnamefont {Haroche}},\ }\href {\doibase 10.1038/nature10376} {\bibfield  {journal} {\bibinfo  {journal} {Nature}\ }\textbf {\bibinfo {volume} {477}},\ \bibinfo {pages} {73–77} (\bibinfo {year} {2011})}\BibitemShut {NoStop}%
\bibitem [{\citenamefont {Di\'osi}(2023)}]{PhysRevA.107.062206}%
  \BibitemOpen
  \bibfield  {author} {\bibinfo {author} {\bibfnamefont {L.}~\bibnamefont {Di\'osi}},\ }\href {\doibase 10.1103/PhysRevA.107.062206} {\bibfield  {journal} {\bibinfo  {journal} {Phys. Rev. A}\ }\textbf {\bibinfo {volume} {107}},\ \bibinfo {pages} {062206} (\bibinfo {year} {2023})}\BibitemShut {NoStop}%
\bibitem [{\citenamefont {Scovil}\ and\ \citenamefont {Schulz-DuBois}(1959)}]{PhysRevLett.2.262}%
  \BibitemOpen
  \bibfield  {author} {\bibinfo {author} {\bibfnamefont {H.~E.~D.}\ \bibnamefont {Scovil}}\ and\ \bibinfo {author} {\bibfnamefont {E.~O.}\ \bibnamefont {Schulz-DuBois}},\ }\href {\doibase 10.1103/PhysRevLett.2.262} {\bibfield  {journal} {\bibinfo  {journal} {Phys. Rev. Lett.}\ }\textbf {\bibinfo {volume} {2}},\ \bibinfo {pages} {262} (\bibinfo {year} {1959})}\BibitemShut {NoStop}%
\bibitem [{\citenamefont {Kalaee}\ \emph {et~al.}(2021)\citenamefont {Kalaee}, \citenamefont {Wacker},\ and\ \citenamefont {Potts}}]{PhysRevE.104.L012103}%
  \BibitemOpen
  \bibfield  {author} {\bibinfo {author} {\bibfnamefont {A.~A.~S.}\ \bibnamefont {Kalaee}}, \bibinfo {author} {\bibfnamefont {A.}~\bibnamefont {Wacker}}, \ and\ \bibinfo {author} {\bibfnamefont {P.~P.}\ \bibnamefont {Potts}},\ }\href {\doibase 10.1103/PhysRevE.104.L012103} {\bibfield  {journal} {\bibinfo  {journal} {Phys. Rev. E}\ }\textbf {\bibinfo {volume} {104}},\ \bibinfo {pages} {L012103} (\bibinfo {year} {2021})}\BibitemShut {NoStop}%
\bibitem [{\citenamefont {Hegde}\ \emph {et~al.}(2024)\citenamefont {Hegde}, \citenamefont {Potts},\ and\ \citenamefont {Landi}}]{TimeResolved}%
  \BibitemOpen
  \bibfield  {author} {\bibinfo {author} {\bibfnamefont {A.~S.}\ \bibnamefont {Hegde}}, \bibinfo {author} {\bibfnamefont {P.~P.}\ \bibnamefont {Potts}}, \ and\ \bibinfo {author} {\bibfnamefont {G.~T.}\ \bibnamefont {Landi}},\ }\href {https://arxiv.org/abs/2408.00694} {\enquote {\bibinfo {title} {Time-resolved stochastic dynamics of quantum thermal machines},}\ } (\bibinfo {year} {2024}),\ \Eprint {http://arxiv.org/abs/2408.00694} {arXiv:2408.00694 [quant-ph]} \BibitemShut {NoStop}%
\bibitem [{\citenamefont {Aamir}\ \emph {et~al.}(2025)\citenamefont {Aamir}, \citenamefont {Jamet~Suria}, \citenamefont {Mar{\'i}n~Guzm{\'a}n}, \citenamefont {Castillo-Moreno}, \citenamefont {Epstein}, \citenamefont {Yunger~Halpern},\ and\ \citenamefont {Gasparinetti}}]{Aamir2025}%
  \BibitemOpen
  \bibfield  {author} {\bibinfo {author} {\bibfnamefont {M.~A.}\ \bibnamefont {Aamir}}, \bibinfo {author} {\bibfnamefont {P.}~\bibnamefont {Jamet~Suria}}, \bibinfo {author} {\bibfnamefont {J.~A.}\ \bibnamefont {Mar{\'i}n~Guzm{\'a}n}}, \bibinfo {author} {\bibfnamefont {C.}~\bibnamefont {Castillo-Moreno}}, \bibinfo {author} {\bibfnamefont {J.~M.}\ \bibnamefont {Epstein}}, \bibinfo {author} {\bibfnamefont {N.}~\bibnamefont {Yunger~Halpern}}, \ and\ \bibinfo {author} {\bibfnamefont {S.}~\bibnamefont {Gasparinetti}},\ }\href {\doibase 10.1038/s41567-024-02708-5} {\bibfield  {journal} {\bibinfo  {journal} {Nature Physics}\ }\textbf {\bibinfo {volume} {21}},\ \bibinfo {pages} {318} (\bibinfo {year} {2025})}\BibitemShut {NoStop}%
\end{thebibliography}%

\appendix

\section{Proof of the jump-based feedback master equation~\eqref{eq: jump FB dyn}}
\label{ap: jump fb dyn}

Let us consider a sequential measurement scheme in which $x_n$ denotes the measurement outcome obtained at time $t_n$. We introduce the \emph{memory function}, denoted by $y_n$, as a general function of the dataset $x_{1:n}\equiv(x_1,\ldots,x_n)$ collected after $n$ detections.
For instance, $y_n$ may correspond to the full dataset $x_{1:n}$, or a compressed summary of it, such as the sample average $y_n = \frac{1}{n}\sum_{i=1}^n x_i$.
The sequential measurement scheme is mathematically described by a set of \emph{instruments} $\{\mathcal{M}_x\}$, consisting of trace–non-increasing maps whose sum $\sum_x \mathcal{M}_x$ is a quantum channel (CPTP map). These superoperators determine both the probability distribution of the outcome $x$ given the state $\rho$, $P(x|\rho) = \mathrm{Tr}[\mathcal{M}_x(\rho)]$, and the corresponding state-update rule conditioned on observing $x$, $\rho \longrightarrow \frac{\mathcal{M}_x(\rho)}{P(x|\rho)}$.

The instruments $\{\mathcal{M}_x\}$ describe the joint effect of the dynamical evolution and the measurement process.
The most general feedback scheme is then introduced by allowing the instruments at time step $t_{n+1}$ to depend on the memory $y_n$, leading to the following update rules
\begin{equation}
\begin{aligned}
    P(x_{n+1} | x_{1:n}) &= \mathrm{Tr}[\mathcal{M}_{x_{n+1}}(y_{n})\rho_{x_{1:n}}],
    \\[0.2cm]
    \rho_{x_{1:n+1}} &= \frac{\mathcal{M}_{x_{n+1}}(y_{n})\rho_{x_{1:n}}}{P(x_{n+1} | x_{1:n})},
\end{aligned}    
\end{equation}
where $\rho_{x_{1:n}}$ denotes the state conditioned on the dataset $x_{1:n}$. 
In other words, the memory $y_n$ is used to modify the measurement process and/or the dynamical evolution at each time step $t_n$.

One can introduce the \emph{memory-resolved state} $\varrho_n(y)$ as
\begin{equation}
    \varrho_n(y) = E[\rho_{x_{1:n}} \delta_{y,y_n}]~,
\end{equation}
where $E[\cdot]$ denotes the average over all possible outcomes $x_{1:n}$, $\delta_{a,b}$ is the Kronecker delta, and $y$ is a possible realization of the stochastic memory $y_n$. 
Let us consider a \emph{causal} memory $y_n$ that updates at each step according to an update function $f_n$, namely $y_{n+1} = f_{n+1}(x_{n+1},y_n)$.
Ref.~\cite{Rosal2025DeterministicFeedback} showed that, for a general feedback protocol based on a causal memory function $y_n = f_n(x_n,y_{n-1})$, the corresponding memory-resolved state $\varrho_n(y)$ evolves according to the deterministic equation
\begin{equation}
    \label{eq: deterministic eq 1}
    \varrho_{n+1}(y) = \sum_{x',y'} \delta_{y, f_{n+1}(x',y')} \mathcal{M}_{x'}(y') \varrho_n(y')~,
\end{equation}
where the sum runs over all possible outcomes $x'$ and all possible realizations $y'$ of $y_n$.

For quantum jump detections, the instruments are given by~\cite{Tutorial}
\begin{align}
    \label{eq: no-jump instrument}
    \mathcal{M}_0\rho &= (1+\delta t \mathcal{L}_0 ) \rho, 
    \\[0.2cm]
    \label{eq: jump instrument}
    \mathcal{M}_k\rho &= \delta t \mathcal{J}_k \rho,
\end{align}
where $\mathcal{M}_0$ corresponds to a no-jump detection and $\mathcal{M}_k$ to a jump detected in channel $k\in\Sigma$, with $\Sigma$ denoting the set of experimentally monitored jump channels. 
The jump superoperator $\mathcal{J}_k$ and the no-jump Liouvillian $\mathcal{L}_0$ are defined in Eqs.~\eqref{eq: jump channel superoperator} and \eqref{eq: no jump liovillian}, respectively. 
We consider a sequential monitoring of quantum jumps, where detections are performed at regular times $t_n = n\,\delta t$ for $n=1,2,\ldots$, with $\delta t>0$ an infinitesimal time interval. 
Thus, the detection outcome $x_n$ is either $x_n = 0$ for a no-jump event or $x_n = k\in\Sigma$ for a jump in channel $k$. 
We then define the \emph{jump memory} $k_n$ as
\begin{equation}
\label{eq: jump_filter}
k_n = x_n + k_{n-1}\delta_{x_n, 0} ~.
\end{equation}
We consider an initial condition $k_0\in\Sigma$. If a jump is detected at time $t_n$ (i.e., $x_n = k$), the jump memory is updated to record this event, $k_n = k$. For a no-jump detection ($x_n = 0$), the memory retains its previous value, $k_n = k_{n-1}$. Hence, $k_n$ always stores the channel corresponding to the most recently detected jump. 
Furthermore, $k_n$ defines a causal memory with update function $f_n(x,q) = x + q\,\delta_{x,0}$.

Therefore, a general jump-based feedback consists of allowing the instruments to depend on the jump memory $k_n$ at each step. In this case, at time $t_{n+1} = (n+1)\,\delta t$ one has $\mathcal{M}_0(k_n)\rho = (\mathrm{1} + \delta t\,\mathcal{L}_0(k_n))\rho$ and $\mathcal{M}_q(k_n)\rho = \delta t\,\mathcal{J}_q(k_n)\rho$.
Considering these instruments together with the jump memory, Eq.~\eqref{eq: deterministic eq 1} yields 
\begin{eqnarray}
       \varrho_{n+1}(k) &= &\sum_{x = 0,~ x\in\Sigma}\sum_{q\in\Sigma} \delta_{k,x+q~\delta_{x,0}} \mathcal{M}_{x}(q) \varrho_n(q)~,\\
       & = & \sum_{q\in\Sigma} \delta_{k,q} \mathcal{M}_{0}(q) \varrho_n(q)\\
       && +\sum_{x\in\Sigma}\sum_{q\in\Sigma} \delta_{k,x} \mathcal{M}_{x}(q) \varrho_n(q)~,
\end{eqnarray}
where we have first evaluated the sum over $x$. Next, using the Kronecker deltas to perform the remaining sum, we obtain
\begin{eqnarray}
       \varrho_{n+1}(k)  =  \mathcal{M}_{0}(k) \varrho_n(k)+\sum_{q\in\Sigma} \mathcal{M}_{k}(q) \varrho_n(q)~,
\end{eqnarray}
and using the definition of the quantum jump instruments, one has 
\begin{equation}
    \varrho_{n+1}(k)  =  (\mathrm{1}+ \delta t \mathcal{L}_0(k)) \varrho_n(k)+\delta t\sum_{q\in\Sigma} \mathcal{J}_{k}(q) \varrho_n(q)~.
\end{equation}
In the continuous monitoring limit $\delta t\to0$, one has $[\varrho_{n+1}(k) - \varrho_n(k)]/\delta t \to \partial_t \varrho_t(k)$, and $\varrho_{n}(k) = \varrho_{t = n\delta t} (k) \to \varrho_t(k)$, and we can write
\begin{eqnarray}
\label{eq_app: jump-based feedback dyn}
       \partial_t \varrho_t(k) =   \mathcal{L}_0(k) \varrho_t(k)+\sum_{q\in\Sigma} \mathcal{J}_{k}(q) \varrho_t(q)~.
\end{eqnarray}
Finally, by using Eq.~\eqref{eq: no-jump liovilian and hamiltonian}, one completes the proof of Eq.~\eqref{eq: jump FB dyn}.

\section{Proof of Result~\eqref{result1: Joint fb dynamics}}
\label{ap: proof of the joint FB dyn}
The composite state that describes the hybrid classical-quantum system is given by
\begin{eqnarray}
    \rho_{\text{sm}}(t) &=& \sum_{k\in\Sigma} P_t(k) ~\rho_t(k) \otimes \ket{k}\bra{k}\\
    &=& \sum_{k\in\Sigma}\varrho_t(k) \otimes \ket{k}\bra{k}~,
\end{eqnarray}
where we used $\rho_t(k) \equiv \varrho_t(k)/P_t(k)$.
Differentiating both sides with respect to time and substituting $\partial_t \varrho_t(k)$ from Eq.~\eqref{eq: jump FB dyn}, one finds
\begin{eqnarray}
\label{eq: proof of result 1}
    \partial_t \rho_{\text{sm}}(t) &=& (-i)\sum_k [H(k),\varrho_t(k)] \otimes\ket{k}\bra{k} \nonumber\\
    && + \sum_k \left( - \frac{1}{2} \sum_q \{L_q^\dagger(k) L_q(k), \varrho_t(k)\}\right) \otimes\ket{k}\bra{k} \nonumber\\
    &&+ \sum_k \left( \sum_q L_k(q) \varrho_t(q) L_k^\dagger(q)\right)\otimes\ket{k}\bra{k}~.
\end{eqnarray}

The first term can be written as
\begin{eqnarray}
    &\sum_k [H(k),\varrho_t(k)] \otimes\ket{k}\bra{k} =\\
   & \sum_k \left(H(k)\varrho_t(k) - \varrho_t(k) H(k)\right) \otimes\ket{k}\bra{k}\nonumber= \\
  & \left( \sum_q H(q)\otimes\ket{q}\bra{q}\right) \left(\sum_k \varrho_t(k) \otimes\ket{k}\bra{k}\right)\nonumber\\
  & - \left(\sum_k \varrho_t(k) \otimes\ket{k}\bra{k}\right) \left( \sum_q H(q)\otimes\ket{q}\bra{q}\right)\nonumber\\
  & = [\mathbb{H},\rho_{\text{sm}}(t)]\nonumber~,
\end{eqnarray}
where we define $\mathbb{H} \equiv \sum_{q\in\Sigma} H(q)\otimes\ket{q}\bra{q}$. 
Using a similar algebra, one can see that
\begin{eqnarray}
   & \sum_{k,q} \mathcal{D}[\mathbb{L}_{k,q}] \rho_{\text{sm}}(t) =\sum_k \left(\sum_q L_k(q) \varrho_t(q) L_k^\dagger(q)\right) \otimes\ket{k}\bra{k}\nonumber \\
   &  \sum_k \left(-\frac{1}{2}\sum_q \{L_q^\dagger(k) L_q(k),\varrho_t(k)\}\right)  \otimes\ket{k}\bra{k}  ~,
\end{eqnarray}
where $\mathcal{D}[L]\rho = L\rho L^\dagger - \tfrac{1}{2}\{L^\dagger L,\rho\}$ denotes the Lindblad dissipator, and $\mathbb{L}_{k,q} \equiv L_k(q)\otimes\ket{k}\bra{q}$.  
This operator reproduces the last two terms in Eq.~\eqref{eq: proof of result 1}.
Hence, one has
\begin{equation}
    \partial_t\rho_{\text{sm}}(t) = -i [\mathbb{H},\rho_{\text{sm}}(t)] + \sum_{k,q \in\Sigma} \mathcal{D}[\mathbb{L}_{k,q}] \rho_{\text{sm}}(t)~,
\end{equation}
as described by Result~\eqref{result1: Joint fb dynamics}.

\section{Analytical expressions for the three-level maser}
\label{ap: expressions maser}

\subsection{Feedback steady-state}
The feedback steady-state can be obtained by solving the coupled equations for $\partial_t \varrho_t(k) = 0$, or, equivalently, by determining the eigenvector of the generator satisfying $\mathbb{L}\rho_{\text{sm}}(t)=0$, where $\mathbb{L}$ is defined in Result~\eqref{result1: Joint fb dynamics}. 
For the feedback protocol implemented in the three-level maser, the stationary maser populations are given by
\begin{eqnarray}
\bra{0}\bar{\rho}_{ss}\ket{0} &=& \big(\eta +  \bar{n}_r \bar{n}_l (1 + \bar{n}_l)\bar{n}\, p^2\big)/\xi~,\\
\bra{1}\bar{\rho}_{ss}\ket{1} &=&\big((1 + \bar{n}_r)\, \bar{n}_l \big(4 + \bar{n}_l \bar{n} p^2 \big)\big)/\xi ~,\\
\bra{2}\bar{\rho}_{ss}\ket{2} &=&\big( \bar{n}_l \bar{n}\,\big(4 + \bar{n}_r \bar{n}_l\, p^2\big)\big)/\xi ~.
\end{eqnarray}
where
\begin{eqnarray}
    \xi &\equiv& 4\big(\bar{n}_r + 4 \bar{n}_r \bar{n}_l + \bar{n}_l(3 + 2 \bar{n}_l)\big)\nonumber\\
    &&+ \bar{n}_l (\bar{n}_r + \bar{n}_l)\big(\bar{n}_r + \bar{n}_l + 3 \bar{n}_r \bar{n}_l\big)\, p^2~, \\
    \eta &\equiv& 4\big(\bar{n}_r + 2 \bar{n}_r \bar{n}_l + \bar{n}_l(2 + \bar{n}_l)\big)~,\\
    \bar{n}&\equiv&\bar{n}_l + \bar{n}_r~,
\end{eqnarray}
and $p \equiv \gamma/\lambda$ represents the competition between the thermal coupling and external drive strength.
Note that both the classical and quantum feedback protocols yield the same stationary populations, whereas the classical protocol produces a density matrix that is always diagonal (i.e., it contains no coherences).

\subsection{Power}
The counting observable considered in the application of the three-level maser consists of the stochastic work defined in Eq.~\eqref{eq: stochastic work of the three level maser}, thereby fixing the general weights $\nu_k$ in Eq.~\eqref{eq: stoc charge int}. Hence, the average current defined in Eq.~\eqref{eq: average current} corresponds to the power delivered by the maser over the external drive, and it is given directly by Eq.~\eqref{eq: average current}. 
Both classical and quantum masers have the same populations, and consequently exhibit the same currents. 
In the no-feedback case, where the external drive is always on, the average power of the maser in the steady-state, denoted by $P_{\text{ss}}^{0}$, is given by
\begin{equation}
\label{eq: power without FB}
    P_{\text{ss}}^{0} = \frac{4(\bar{n}_l-\bar{n}_r)}{4(4 + 3\bar{n}_r + 3\bar{n}_l) + \phi\, p^2}~,
\end{equation}
where we considered that both baths have the same coupling strength to the system, $\gamma_l = \gamma_r \equiv \gamma$, and $\phi \equiv (\bar{n}_r + \bar{n}_l)(\bar{n}_r + \bar{n}_l + 3\bar{n}_r \bar{n}_l)$. 

Note that the sign of $P_{\text{ss}}^{0}$ depends on the difference $\bar{n}_l - \bar{n}_r$. 
A positive power $P_{\text{ss}}^{0} > 0$ corresponds to positive work, on average, performed by the maser on the drive, thereby indicating operation in the engine regime. 
In contrast, $P_{\text{ss}}^{0}<0$ indicates that work is performed by the drive on the maser, characterizing a refrigeration regime. 
Since $\bar{n}_\alpha = \bigl(e^{\omega_\alpha/T_\alpha}-1\bigr)^{-1}$ for $\alpha = l,r$, one finds that $P_{\text{ss}}^{0}>0$ only when $\omega_l/T_l > \omega_r/T_r$.

On the other hand, the average power $P_{\text{ss}}$ in the feedback steady-state is given by
\begin{equation}
    \label{eq: power with FB}
    P_{\text{ss}} =
\frac{4(1 + \bar{n}_r)\bar{n}_l^2}{4\big(\bar{n}_r + 4\bar{n}_r \bar{n}_l + \bar{n}_l(3 + 2\bar{n}_l)\big) + \bar{n}_l\phi\, p^2}~,
\end{equation}
where we also considered  $\gamma_l = \gamma_r \equiv \gamma$.
In this case, the power $P_{\rm ss}$ is always positive, and the maser operates exclusively as an engine.
Equations~\eqref{eq: power without FB} and \eqref{eq: power with FB} are displayed in Fig.~\ref{fig: diagram and plots} (a).

\end{document}